\documentclass[final,5p,twocolumn]{elsarticle}

\usepackage{graphicx}
\usepackage{dcolumn}
\usepackage{bm}
\usepackage{color}

\usepackage{amssymb}
\usepackage{epstopdf}
\usepackage{amstext}
\begin{document}

\begin{frontmatter}

\title{Quantum deformation of quantum cosmology:\\ {\small A framework to discuss the cosmological constant problem}}
\author{S. Jalalzadeh$^{1}$} \ead{shahram.jalalzadeh@unila.edu.br}
\author{A.J.S. Capistrano$^{1,2}$ }\ead{abraao.capistrano@unila.edu.br}
\author{P.V. Moniz$^{3,4}$}\ead{pmoniz@ubi.pt}
\address{ $^{1}$Federal University of Latin-America Integration, Itaipu Technological Park, PO box 2123, Foz do Igua\c{c}u-PR,  85867-670, Brazil}
\address{$^{2}$ Casimiro Montenegro Filho Astronomy Center, Itaipu Technological Park, 85867-900, Foz do Iguassu-PR, Brazil}
\address{$^{3}$ Centro de Matem\'{a}tica e Aplica\c{c}\~{o}es- UBI, Covilh\~{a}, Portugal}
\address{$^{4}$ Departmento de F\'{\i}sica , Universidade da Beira Interior, 6200 Covilh\~{a}, Portugal}
\date{\today}
\begin{abstract}

 We endorse the context that the cosmological constant problem is a quantum cosmology
issue. Therefore, in this paper we investigate the $q$-deformed Wheeler-DeWitt equation  of a spatially closed
homogeneous and isotropic Universe in the presence of a conformally coupled scalar field. Specifically, the quantum deformed Universe
is a quantized  minisuperspace model constructed from quantum
Heisenberg-Weyl $\mathcal U_q(h_4)$ and $\mathcal U_q(su(1, 1))$  groups. These intrinsic mathematical features allow to establish
that ($i$) the
scale factor, the scalar field and corresponding momenta are  quantized   and ($ii$) the phase space has a non-equidistance lattice structure. On the other hand, such quantum group structure provides us a new framework to discuss the cosmological constant problem.  Subsequently, we show that a ultraviolet cutoff can be obtained at $10^{-3} eV$, i.e.,  at a scale  much larger than the expected
Planck scale. In addition,  an
infrared cutoff, at the size of the observed  Universe, emerges from within such  quantum
deformation of Universe.   In other words,  the spectrum of the scale factor is upper bounded.
Moreover, we show that the emerged cosmological horizon is a quantum sphere $S^2_q$ or, alternatively, a fuzzy sphere $S^2_F$ which explicitly exhibits features of the holographic principle. The corresponding number of fundamental cells  equals  the dimension of the Hilbert space and hence,  the cosmological constant can be presented as a consequence of the quantum deformation of the FLRW minisuperspace.
\end{abstract}
\end{frontmatter}

\section{Introduction}

{Since the mid-1980s, astrophysicists have been compiling evidence --
such as
cosmic microwave background observations, the supernova type Ia data and large scale structure -- that the late time Universe is accelerating. The simplest candidate
to explain this acceleration, within the framework of general relativity
(GR),  is a positive cosmological constant (CC). Many theoretical physicists were reluctant
to consider the CC as a {\it bona fide} explanation regarding the mentioned acceleration, because the natural
predicted value for the  CC
from particle physics is $\rho_\Lambda\simeq M_\text{P}^4\simeq(10^{18}GeV)^4$,
which has a enormous discrepancy with the astronomical bound for CC, $\rho_\Lambda\simeq(10^{-3}eV)^4$--some
$10^{122}$ times too small.  In other words, from an effective field theory (EFT) point
of view, the CC is the zero point energy with a UV cutoff scale, for example the Planck scale or the supersymmetry breaking scale, and on the other hand,  from the
cosmological point of view,  the CC is an IR scale problem
and affecting the large scale structure
of the Universe, when we investigate the whole Universe.
Hence, the CC problem seems to violate our prejudice about decoupling
 UV and IR scales, which underlies the
use of EFT. The CC, being interpretable both
as the zero point energy and as the scale of the observed Universe, goes against the
notion of local quantum fields and suggests a mixing between local UV and global IR physics. In this direction, some physicists believe that CC problem is essentially
a quantum gravity and quantum cosmology problem \cite{Banks}.
A candidate theory for quantum gravity  must provide a classical continuum
spacetime geometry at  macroscopic scales with a global IR cutoff (CC) but also involving quantum corrections at the local UV scale.}

The quantum spacetime hypothesis asserts that the classical continuum should break down at the Planck scale. As we know, the Planck length provides a natural length
unit involving  gravitational and quantum features. It defines a distance
scale at which quantum corrections to GR are expected to be significant.
It is also commonly thought to provide a momentum cutoff, rendering finite otherwise  divergent
particle self-energies and enabling gravity to play the role of a universal
regulator to other fundamental interactions.

It is well known that  noncommutative (NC) geometry \cite{Connes1}  provides an approach to deal with possible properties of Planck scale physics \cite{Connes2}. 
In fact, the idea of the quantization of a spacetime manifold, as well as  phase space symplectic
manifolds, using noncommuting
coordinates $\hat x^\mu$, is an old one \cite{NC8}. In this
context, the quantum spacetime of Hartland Snyder type
introduces\footnote{One of the most interesting applications of this type of noncommutativity
to quantum field theory, is that concerning a description of Yang-Mills instantons
in NC spacetime; In these NC spaces, instantons acquire an effective size proportional
to the noncommutativity parameter $\theta$. As a consequence, the moduli space of NC instantons
no longer has the singularities corresponding to small instantons \cite{Instanton}.}
\begin{eqnarray}\label{in1}
[\hat x^\mu,\hat x^\nu]=i\theta^{\mu\nu},
\end{eqnarray}
where the $\theta_{\mu\nu}$ generates the Lorentz group \cite{Snyder}.
The corresponding NC gravity has been considered in various models (see for
example \cite{N1,N2}). In particular in Ref.\cite{N2} a NC Einstein
gravity is constructed by using the Seiberg-Witten
map and gauging the NC ISO(3,1) group. Also, the consequence of
Snyder type of NC has been studied in the context of homogeneous cosmologies
for various minisuperspace  models
\cite{NC}.

Moreover, fuzzy NC  models proposed by  't Hooft \cite{Fuzzy}  have
\begin{eqnarray}\label{f11}
[\hat x^\mu,\hat x^\nu]=2i\lambda\varepsilon_{\mu\nu\rho}\hat x^\rho.
\end{eqnarray}
{ It is, in fact, a toy model of quantum gravity in a $3D$ manifold with a Euclidean signature.}
A subsequent development, including  quantum differential calculus and an action of a certain ``quantum double'' quantum group as NC Euclidean group of motions, was proposed by Majid and E. Batista \cite{Bat}.

Furthermore, the Majid-Ruegg bicrossproduct model spacetimes \cite{Bi}  bear
\begin{eqnarray}\label{f12}
[\hat x^i,\hat x^j]=0,\,\,[\hat x^i,\hat t]=i\lambda\hat x^i,\,\,i,j=1,2,3,
\end{eqnarray}
with corresponding deformed Poincar\'e group and convey a physically testable prediction of a variable speed of light.

{ In Ref.\cite{Cam} the authors showed that in Snyder type NC spacetimes defined in (\ref{in1})   there is a minimum distance but no minimum
area. This suggests that we need to be careful about  the possible emergence of
minimum area  at the Planck scale \cite{Cam} in this
kind of NC spacetimes. }

{Let us also mention a specific category of NC spacetimes featuring} anisotropic or q-deformed manifolds \cite{Q}, motivated from quantum groups theory,  given by
\begin{eqnarray}\label{in2}
\hat x^\mu\hat x^\nu=q\hat x^\nu\hat x^\mu.
\end{eqnarray}
 This model was developed independently in
\cite{Wess}  and also by Majid and coworkers, in a series of papers on braided matrices \cite{Ma}.

In general, quantum groups give us symmetries which are richer than the classical Lie algebras, which are contained in the former as a special case.  It is therefore possible that
quantum groups\footnote{ Quantum groups and algebras emerged from studies on quantum integrable models using the quantum inverse scattering
method \cite{Fa} and led to certain deformations of
classical matrix groups and the corresponding Lie algebras. The original main reason for the great significance of  quantum groups was that
they are related to the so called quantum Yang-Baxter equation \cite{Chari} which plays a major role in quantum integrable systems,  conformal
field theory \cite{Oh}, solvable lattice models \cite{Omar}, knot theory
\cite{Kaf},  topological quantum computation \cite{Top}. Phenomenological applications of quantum groups in nuclear \cite{Den} and molecular spectroscopy
\cite{Cha} lead to significant results showing that the vibrational-rotational spectra of nuclei and molecules can be fit into schemes in which the number of phenomenological
deformation parameters required are very much fewer than the number of traditional
phenomenological parameters. } can turn out to be suitable for describing symmetries of physical systems which
are outside the realm of Lie algebras \cite{Den}.
Therefore,
{ the $q$-deformed models spacetimes are applicable
for any physical manifold. In addition, the great advantage of $q$-deformed models is that
the corresponding Hilbert space is finite dimensional when $q$ is root of
unity \cite{Chaichian}. This suggests that the use of quantum groups when deformation parameter  is root of unity is a powerful tool
to build models with a finite number of states, aiming at  applications in quantum gravity and quantum cosmology that are consistent with the assumption of the holographic
principle\footnote{{ Any consistent theory of quantum gravity requires dramatically new ideas like
the holographic principle \cite{Holo} which have recently attracted increasing attention.} Perhaps the most remarkable
aspect of this principle is that the entropy is finite,  suggesting a finite dimensional Hilbert space of excitations describing the interior of a region bounded by a surface of area. This is also a feature found  in loop quantum gravity, where quantum groups at roots of unit and finite number
of states naturally enter the formalism \cite{q-0}.} and a UV/IR mixing to solve CC problem.}

Quantum groups may appear in  $q$-versions of gravity in various situations.  In Ref.\cite{Fin}
Finkelstein  constructed a $q$-deformation of GR by replacing the Lorentz group by the quantum Lorentz group.
Furthermore, in \cite{Cas} a
$q$-gravity is constructed by ``gauging'' the
quantum analogue of a Poincar\'e algebra. {They
also occur in Hamiltonian quantization formalisms for $(2+1)$-gravity such as combinatorial quantization \cite{Com}. In spin foam models, quantum groups assure a cut off against IR divergences  \cite{Spin, Han}. In addition to this, quantum group symmetries also arise  in NC geometry models such as $\kappa$-Poincar\'e models \cite{Po} and related doubly special relativity theories \cite{SR}.}

  Generally speaking there are three
possibilities \cite{Ya} for a $q$-deformation of a classical and $\hbar$-deformed
(usual quantum mechanical) systems:
($i$)- The spacetime manifold is standard commutative and  the variables
in a field theory take values in a
quantum group.
($ii$)- Classical mechanics or field theory on $q$-deformed
spacetime, i.e. field dynamical variables are
defined on quantum spacetime and take values in
the usual  algebras.
($iii$)- Variables in field
theory take values in a quantum manifold  and spacetime
is also noncommutative. In the present paper we will use the first route. More specifically, we consider the quantum deformation of
the phase space variables from a spatially closed
homogeneous and isotropic universe minisuperspace, with $q$ being a  root of unit.

 This paper is organized as follows. In Sec. II, we present the model
that assists in our investigation. In
Sec.III, the corresponding WDW equation and   boundary conditions of model are extracted. In Sec.IV,
 we describe the hidden symmetries and
 corresponding quantum
groups  that are deformations of the
enveloping algebras of Heisenberg-Weyl and $su(1,1)$ Lie algebras. {We  obtain the eigenvectors and discrete eigenvalues of scale factor, scalar field
and the corresponding momenta.
We show that the scale factor and the corresponding momenta are non-singular. In section
V we explains  that the finite dimensional Hilbert space of
the model leads to inherent UV/IR mixing. Subsequently, we describe that the emerged
horizon is a fuzzy sphere and subsequently the emerged CC  is consistence with observations. { Also, we show that how the quantum deformation of model may solve the coincidence problem.}
}  In Sec. VI, we present our concluding remarks.  Hereon, we use natural units, $c=\hbar=1$.

\section{Classical setting}

  The action of GR with a conformally coupled scalar field is
\begin{eqnarray}\label{1-1}
\begin{array}{cc}
S=\frac{M^2_\text{p}}{2}\int\sqrt{-g}Rd^4x+\\
\\+\frac{1}{2}\int\sqrt{-g}\left(g^{\mu\nu}\nabla_\mu\phi\nabla_\nu\phi-\frac{1}{6}R\phi^2\right)d^4x,
\end{array}
\end{eqnarray}
where $M_\text{P}=1/\sqrt{8\pi G}$ is the reduced Planck mass, $R$ is the scalar
curvature of spacetime manifold $(g,\nabla)$ and $\phi$ is the scalar field. There are various reasons to include a non-minimally
coupled scalar field to the action. The first one which we may consider is that at the quantum level, quantum
corrections to the scalar field theory lead to the non-minimal
coupling in the sense that the scalar field theory in curved spacetime
becomes renormalizable in the case of a non-minimal
coupling \cite{C1}. Furthermore, recent  Planck data \cite{C2}
suggest for the early Universe a stage where a nonminimal
coupling may have had a suitable contribution.

Let us consider the spacetime line element is of the Robertson-Walker form
\begin{eqnarray}\label{1-2}
ds^2=-N^2(t)dt^2+a^2(t)\left(\frac{dr^2}{1-r^2}+r^2d\Omega^2\right),
\end{eqnarray}
where $N(t)$ is the lapse function, $a(t)$ is the scale factor and $d\Omega^2$ is the line element of standard unit 2-sphere. Furthermore, we shall assume the
scalar field shares the symmetry of minisuperspace, so that $\phi=\phi(t)$
only.
By substituting (\ref{1-2}) in action functional (\ref{1-1}), rescaling
the lapse function as $N(t)=12\pi^2M_\text{P}a(t)\tilde N(t)$ and defining the new
variables
\begin{eqnarray}\label{1-3}
x_1=a,\,\,\,\,\,\,\,\,\,x_2=\frac{1}{\sqrt{6}M_\text{P}}a\phi,
\end{eqnarray}
the action (\ref{1-1}) reduces to the following action in two-dimensional
minisuperspace $x_i=\{x_1,x_2\}$
\begin{eqnarray}\label{1-4}
\begin{array}{cc}
S=\\-\int\left\{\frac{M_\text{P}}{2\tilde N}\left(\dot x_1^2-\dot x_2^2\right)-\frac{1}{2}M_\text{P}\omega^2\tilde
N\left(x_1^2-x_2^2\right)\right\}dt,
\end{array}
\end{eqnarray}
where $\omega=12\pi^2M_\text{P}$. It should be noted
that an action like (\ref{1-4}) also arises in an interior solution of Schwarzschild
black hole \cite{Interior}, varying speed of light cosmology \cite{Light}, Kaluza-Klein \cite{KK} and
multidimensional \cite{MD} cosmological models.

Since the lapse function is not dynamical, the super-Hamiltonian
vanishes
\begin{eqnarray}\label{1-5}
{\mathcal H}=-\frac{1}{2M_\text{P}}\left(\Pi_1^2-\Pi_2^2\right)-\frac{1}{2}M_\text{P}\omega^2\left(x_1^2-x_2^2\right)=0,
\end{eqnarray}
where $\Pi_1=-\frac{M_\text{P}}{\tilde N}\dot x_1$ and $\Pi_2=\frac{M_\text{P}}{\tilde N}\dot x_2$ are the conjugate momenta of $x_1$ and $x_2$ respectively.
Classical solutions of field equations at the gauge $\tilde N=1$ are given
by
\begin{equation}
    \label{1-6}
x_1=B \sin(\omega t),\,\,\,\,x_2=\pm B\sin(\omega t+\theta),
\end{equation}
where $B$ represents the maximum value of the scale factor. In addition, the relation between comoving
time, $\eta$, and conformal time, $t$, is given by $\eta=B(1-\cos(\omega t))$. Therefore, $B$ also represents the value of comoving time when the
scale factor is maximum, $a_\text{Max}(\eta)=B$. Note that $B$
can take any  positive value.

\section{Canonical quantization}

{Traditionally, the question of boundary conditions in cosmology has been split into two parts. The first part is that  the spatial sections of a spatially homogeneous and isotropic Universe have a boundary. The second part is what are the initial
conditions for the corresponding quantum cosmology. Einstein preferred spatially
closed cosmological models because he believed this eliminated
the first of these questions \cite{Ei}.
Moreover, Linde \cite{Linde}  showed that it is possible
to produce
inflationary models which result in closed
universes in which the universe is ``created from nothing''
\cite{TZ}. This is possible because a closed
universe has zero total energy, just as it has zero
total momentum and total charge \cite{Lan}. Various authors have tried to address the second question which is about the boundary conditions in quantum
cosmology. Two leading well known
lines (for closed universes) with a false vacuum energy are the no-boundary proposal \cite{Haw} and the tunneling proposal \cite{Ver}. Two other
proposals have been used as explicit procedures to deal with the presence
of classical singularities. More precisely, the wave function
should vanish at the classical singularity  (DeWitt or Neumann
boundary condition) \cite{De}, or its derivative with respect to
the scale factor vanishes at the classical singularity (Dirichlet boundary condition) \cite{Mosh}.

 Our physical interpretation of quantum mechanical operators depends on their
Hermicity and self-adjointedness. However, in some cases boundary conditions must be specified in order for Hamiltonians to be Hermitian and self-adjoint. In particular, Hamiltonians with singular potentials or restricted domain
of definition require us to specify how wave functions behave at the boundaries
or at the singularities of potential. Thus, a satisfactory treatment of the WDW equation of our cosmological model requires to
solve the equation in a Hilbert space and the solutions have to be associated
with a self-adjoint operator.  }

In coordinate representation, the canonical quantization of our model
is accomplished by setting $x_i = x_i$ and $\Pi_i = -i\frac{\partial}{\partial x^i}$.  Then,
the Hamiltonian constraint (\ref{1-5}) becomes the WDW equation
\begin{equation}\label{2-1}
-\frac{1}{2M_\text{P}}\left(\partial_1^2-\partial_2^2\right)\Psi(x_i)+\frac{1}{2}M_\text{P}\omega^2(x_1^2-x_2^2)\Psi(x_i)=0.
\end{equation}
Due to the hyperbolic character of (\ref{2-1}), we can separate
the scalar field part from the gravitational sector, i.e., ${\mathcal H}={\mathcal H}_1\oplus {\mathcal H}_2$
where ${\mathcal H}_1$ and ${\mathcal H}_2$ represent the gravitational and scalar field parts
of super-Hamiltonian (\ref{1-5}) respectively. By assuming $\Psi(x_1,x_2)=\Theta(x_1)\Phi(x_2)$,
for the scalar field part $\Phi(x_2)$, with a separation constant $E$, we find
\begin{eqnarray}\label{2-2}
\left(-\frac{1}{2M_\text{P}}\frac{d^2}{dx_2^2}+\frac{1}{2}M_\text{P}\omega^2x_2^2\right)\Phi(x_2)=E\Phi(x_2).
\end{eqnarray}
The solution to the above equation is
\begin{eqnarray}\label{2-3}
\begin{array}{cc}
\Phi_n(x_2)= C_nH_n\left(\sqrt{M_\text{P}\omega}x_2\right)e^{-\frac{M_\text{P}\omega}{2}x_2^2},\\
E_n=\omega(n+\frac{1}{2}),
\end{array}
\end{eqnarray}
where $H_n$ are the Hermite polynomials.

The domain of definition of scale factor, $x_1=a$, is $\mathbb{R}^+$. Therefore, the gravitational part of WDW equation
\begin{eqnarray}\label{2-4}
\left(-\frac{1}{2M_\text{P}}\frac{d^2}{dx_1^2}+\frac{1}{2}M_\text{P}\omega^2x_1^2\right)\Theta(x_1)=E\Theta(x_1),
\end{eqnarray}
is defined on a dense domain ${\mathcal D}(\mathcal H_1)=C_0^\infty(\mathbb{R}^+)$. The operator ${\mathcal H}_1:=-\frac{1}{2M_\text{P}}\frac{d^2}{dx_1^2}+\frac{1}{2}M_\text{P}\omega^2x_1^2$
within the square-integrable Hilbert space $L^2({\mathbb R}^+)$ is in the limit point case at $+\infty$ and in the limit circle case at zero,
hence it is not essentially
self-adjoint \cite{St}.  ${\mathcal
H}_1$ is
Hermitian if
\begin{eqnarray}\label{2-5}
\langle\Theta_1|{\mathcal H}_1\Theta_2\rangle=\langle {\mathcal H}_1\Theta_1|\Theta_2\rangle,\,\,\,\,\,\Theta_1,\Theta_2\in\mathcal
D(\mathcal H_1).
\end{eqnarray}
{Since $\mathcal H_1$ is not singular, such that all subtleties related to Hermiticity and self-adjointness are associated entirely with the behavior at $x_2=0$,} this is the case if
\begin{eqnarray}\label{2-51}
\lim_{x_1\rightarrow0^+}\left(\frac{d\Theta_1^*}{dx_1}\Theta_2-\Theta^*_1\frac{d\Theta_2}{dx_1}\right)=0.
\end{eqnarray}
It can be shown \cite{Neumann} that to ensure the validity of this condition it is necessary and sufficient
for the domain of $\mathcal H_1$ to be restricted to those wave functions that satisfy the
Robin boundary condition
\begin{eqnarray}\label{2-6}
\frac{d\Theta_1}{dx_1}(0^+)+\gamma\Theta_1(0^+)=0,
\end{eqnarray}
where $\gamma$ is an arbitrary real constant which has the  dimension of inverse of length.  {The parameter $\gamma$ thus characterizes
a 1-parameter family of self-adjoint extensions of the $\mathcal H_1$ on the half-line.

There is a peculiar  a difficulty here with these extensions, which is that each extension leads to
a different physics and the problem is not just of technical nature. Nevertheless, due to the existence of the conformal scalar field  we will show that the scalar field part solves the problem.}
The general square-integrable solution of Eq.(\ref{2-4}) with boundary condition
(\ref{2-6}) is given by
\begin{eqnarray}\label{2-7}
\begin{array}{cc}
\Theta(x_1)=
\frac{\sqrt{\pi}e^{-\frac{1}{2}M_\text{P}x_1^2}}{2^{\frac{1}{4}-\frac{E}{2\omega}}\Gamma(\frac{3}{4}-\frac{E}{2\omega})}{_1F_1}(\frac{1}{4}-\frac{E}{2\omega};\frac{1}{2};\frac{M_\text{P}\omega}{2}x_1^2)\\
-\frac{\sqrt{\pi M_\text{P}\omega}x_1e^{-\frac{1}{2}M_\text{P}x_1^2}}{2^{-\frac{3}{4}-\frac{E}{2\omega}}\Gamma(\frac{1}{4}-\frac{E}{2\omega})}{_1F_1}(\frac{3}{4}-\frac{E}{2\omega};\frac{3}{2};\frac{M_\text{P}\omega}{2}x_1^2),
\end{array}
\end{eqnarray}
where $_1F_1(\alpha;\beta;x_1)$ denotes confluent hypergeometric
function. Making use of the properties, $_1F_1(\alpha;\beta;0)=1$ and $\frac{d\,\,_1F_1}{dx}(\alpha;\beta;x)=\frac{\alpha}{\beta}\,\,_1F_1(\alpha+1;\beta+1;x)$,
 we can rewrite the boundary condition (\ref{2-6}) as
 \begin{eqnarray}\label{2-8}
 \gamma=2\sqrt{M_\text{P}\omega}\frac{\Gamma(\frac{3}{4}-\frac{E}{2\omega})}{\Gamma(\frac{1}{4}-\frac{E}{2\omega})}.
 \end{eqnarray}

On the other hand, the scalar field part of WDW equation, (\ref{2-3}), gives
$\frac{E}{\omega}=n+\frac{1}{2}$. Inserting this result into (\ref{2-8}) gives
\begin{eqnarray}\label{2-9}
\gamma=2\sqrt{M_\text{P}\omega}\frac{\Gamma(\frac{1-n}{2})}{\Gamma(-\frac{n}{2})},\,\,\,\,n=0,1,2,...\,.
\end{eqnarray}
For odd values of $n$, this equation fixes the length parameter as $\gamma=+\infty$.
In addition, for even values of $n$, we find $\gamma=0$. Therefore, the scalar field
part of the WDW equation restricts  the self-adjoint extension of gravitational
part to the self-adjoint extension operator with Dirichlet boundary condition
\cite{Mosh}
\begin{eqnarray}\label{2-10}
\frac{d\Theta(x_1)}{dx_1}|_{x_1\rightarrow0^+}=0,
\end{eqnarray}
where the spectrum coincide with the spectrum of the odd parity eigenfunctions
of the harmonic oscillator, or by DeWitt boundary (or Neumann boundary) condition
\cite{De}
 \begin{eqnarray}\label{2-11}
\Theta(x_1)|_{x_1\rightarrow0^+}=0,
\end{eqnarray}
with eigenvalues $E_n=\omega(2n+\frac{1}{2})$, which coincides with the even parity sector of the harmonic oscillator spectrum.

\section{Quantum deformation of quantum cosmology}

{ Two basic concepts both in classical and quantum systems are states
and observables. In classical
mechanics states are points of a phase space manifold, $\Gamma$, and observables
(physical quantities) are smooth functions $f\in C^{\infty}(\Gamma)$.  Every
state determines the value of the observables on that state and covertly
any state is uniquely determined by the values of all observables on it \cite{Franco}. In  quantum mechanics states are one-dimensional subspaces of a Hilbert space
 and observables are operators in Hilbert space.
By the above duality relation between states and observables in both classical
and quantum cases, the relation between classical and quantum mechanics is easier to understand
in terms of an algebra of observables: observables
form an associative algebra which is commutative (abelian) in the classical mechanics and NC (non-abelian)
in the corresponding quantum system. In this regard,  quantization (in quantum mechanics where noncommutativity controlled by $\hbar$) amounts to  replace the
commutative algebras by NC ones \cite{Derin} and all related fundamental mathematical concepts are expressed in such a way that it does
not require commutativity of the algebra. In this manner we may arrive at the concept of
NC geometry: the usual (algebraic) geometry is the study of commutative algebras and
NC (algebraic) geometry is the study of NC algebras. In this regards, NC Hopf algebras are like non-abelian groups. 

Hopf structures in ordinary Lie groups and Lie
algebras.  If  $F(G)$  denotes  the set of differentiable functions from a Lie group $G$ into the complex numbers $\mathbb{C}$, then the algebraic structure is given by the
usual pointwise sum and product of functions and the unit of algebra $I\in
F(G)$ is the constant
function $I(g) = 1$, $\forall g\in G$. Using the group structure of $G$ we can introduce on $F(G)$ three other
linear operations, the coproduct $\Delta$, the counit $\varepsilon$ and the antipode (or coinverse) $S$:
\begin{eqnarray}\label{Ho}
\begin{array}{cc}
\Delta(f)(g_1,g_2)=f(g_1g_2),\\\Delta:F(G)\rightarrow F(G)\otimes F(G);\\
\varepsilon(f)=f(e),\,\,\,\,\,\,\,\,\,\, \varepsilon: F(g)\rightarrow\mathbb{C};\\
(Sf)(g)=f(g^{-1}),\,\,\,\,\,\,\,\,S:F(G)\rightarrow F(G),
\end{array}
\end{eqnarray}
where $e$ is the unit of $G$ and $g_1,g_2\in G$. Algebra $F(G)$ equipped with the linear maps $\Delta$, $\varepsilon$ and $S$ is called
the Hopf algebra \cite{Alg}. Coproducts are
commonly used in the familiar addition of  momentum, angular momentum
and of other so-called primitive operators in quantum mechanics.
Since additivity of observables  is an essential requirement, the coproduct, and therefore
the Hopf algebra structure, appears to provide an essential algebraic tool
in quantum mechanics. Therefore, Hopf algebra is a bialgebra with an antipode.

 Formally, quantum groups are defined to be Hopf algebras which are in general,
NC. From a physical point of view, quantum group includes two basic ideas, namely the
$q$-deformation of an algebraic structure and the notion of a NC comultiplication.
Physicists are familiar with the idea of deformation. For example,  the Poincar\'e group is a deformation of the Galilei group with deformation parameter $c$, which is recovered in the limit $c\rightarrow\infty$ or
quantum mechanics can be considered as a deformation of classical mechanics
with deformation parameter Planck's constant which
is regained in the limit $\hbar\rightarrow0$. In the $q$-deformation of an algebraic structure, usually a
deformation parameter $q$ (a dimensionless parameter) is introduced in which
a commutative algebra becomes  noncommuting
 and in the  $q\rightarrow1$ limit, the original algebraic structure
is recovered. }

\subsection{Quantum deformation of the scalar field}

To construct  the quantum deformation of the scalar field part of the super-Hamiltonian
given in  Eq.(\ref{2-2}),
let us employ the Heisenberg-Weyl algebra of
${\mathcal{H}}_2$, $h_4$. This is a non-semisimple
Lie algebra with four generators $\{A_+,A_-,N,e\}$ that satisfy following commutation
relations
\begin{eqnarray}\label{3-1}
[A_-,A_+]=e,\,\,[N,A_{\pm}]=\pm A_{\pm},
\end{eqnarray}
where $e$ is the central charge and
\begin{eqnarray}\label{3-2}
\begin{array}{cc}
A_+:=\sqrt{\frac{M_\text{P}\omega}{2}}\left(x_2+\frac{1}{M_{P}\omega}\frac{d}{dx_2}\right),\\
A_-:=\sqrt{\frac{M_\text{P}\omega}{2}}\left(x_2-\frac{1}{M_{P}\omega}\frac{d}{dx_2}\right).
\end{array}
\end{eqnarray}
Since $h_4$ is a Lie algebra, its universal enveloping algebra $\mathcal
U(h_4)$ is a Hopf algebra with the usual maps given by
\begin{eqnarray}\label{3-3}
\begin{array}{cc}
\Delta(y)=y\otimes1+1\otimes y,\\
\varepsilon(y)=0,\,\,\,\,\,S(y)=-y,
\end{array}
\end{eqnarray}
where $y\in\{A_+,A_-,N,e\}$, $\Delta$ is the comultiplication of algebra,
$\varepsilon$ denotes its counit and $S$ is antipode. For the unit element of the
algebra the Hopf maps are $\Delta(1)=1$, $\varepsilon(1)=1$ and $S(1)=1$. It
is usual to work with the quotient algebra $\mathcal U'(h_4)=\mathcal U(h_4)/\langle e-1\rangle$,
where the Heisenberg relation is $[A_-,A_+]=1$.

In the Fock space, $\mathcal F_2$, with the basis $\{|n\rangle,\,\,N|n\rangle=n|n\rangle\}$
the pairs of operators $A_\pm$ act in the following form
\begin{eqnarray}\label{3-4}
\begin{array}{cc}
A_+|n\rangle=\sqrt{n+1}|n+1\rangle,\\
A_-|n\rangle=\sqrt{n}|n-1\rangle.
\end{array}
\end{eqnarray}
The quantum Heisenberg-Weyl algebra, $\mathcal U_q'(h_4)$, is the associative
unital \cite{unital} $\mathbb{C}(q)$-algebra with generators $\{\mathcal A_+,\mathcal A_-,q^{\frac{1}{2}\mathcal N}, q^{-\frac{1}{2}\mathcal N}\}$ with the following q-deformed commutation relations \cite{Chaichian}
\begin{eqnarray}\label{3-5}
\begin{array}{ccc}
\mathcal A_-\mathcal A_+-q^{\frac{1}{2}}\mathcal A_+\mathcal A_-=q^{-\frac{1}{2}\mathcal N},\\
\mathcal A_-\mathcal A_+-q^{-\frac{1}{2}}\mathcal A_+\mathcal A_-=q^{\frac{1}{2}\mathcal N},\\

[\mathcal{ N},\mathcal{ A}_\pm]=\pm\mathcal{ A}_\pm.
\end{array}
\end{eqnarray}
Note that in this definition we do not postulate any relation among the generators
of the algebra. We  can show that for this representation, the first two relations
(\ref{3-5}) are actually equivalent to the following relations
\begin{eqnarray}\label{3-6}
\mathcal A_+\mathcal A_-=[\mathcal N],\,\,\,\,\mathcal A_-\mathcal A_+=[\mathcal
N+1],
\end{eqnarray}
where
\begin{eqnarray}\label{3-61}
[y]:=\frac{q^{\frac{1}{2}y}-q^{-\frac{1}{2}y}}{q^{\frac{1}{2}}-q^{-\frac{1}{2}}}.
\end{eqnarray}
The relation $\mathcal A_+\mathcal A_-=[\mathcal N]$ can be compared to the
central element {(it commutes with all generators of algebra)}  of $\mathcal U'(h_4)$, $A_+A_--N$, which acts as zero on
the standard Fock space representation of vacuum. If we define the vacuum
state of the quantum deformed Fock space, $\mathcal F_2(q)$, by
\begin{eqnarray}\label{3-7}
\mathcal A_-|0\rangle=0,\,\,\,\,\,\,\,\,\,q^{\pm\frac{1}{2}\mathcal N}|0\rangle=|0\rangle,
\end{eqnarray}
then we can construct the representation of the $\mathcal U'_q(h_4)$ in the
Fock space spanned by normalized eigenvectors $|n\rangle$
\begin{eqnarray}\label{3-8}
|n\rangle=\frac{1}{\sqrt{[n]!}}{\mathcal A}_+^n|0\rangle,
\end{eqnarray}
where the $q^\frac{1}{2}$-factorial defined by $[n]!:=\prod_{m=1}^n[m]$. The basis $|n\rangle$
defined above is orthonormal due to the identities
\begin{eqnarray}\label{3-9}
\begin{array}{cc}
\mathcal A_-\mathcal A_+^n=q^{\pm\frac{n}{2}}\mathcal A_+^n\mathcal A_-+[n]\mathcal A_+^{n-1}q^{\mp\frac{1}{2}\mathcal
N},\\
\mathcal A_-^n\mathcal A_+=q^{\mp \frac{n}{2}}\mathcal A_+\mathcal A_-^n+[n]\mathcal A_-^{n-1}q^{\pm\frac{1}{2}\mathcal
N}.
\end{array}
\end{eqnarray}
Hence, in the Fock space $\mathcal F_2(q)$ the set up operators act due
\begin{eqnarray}\label{3-10}
\begin{array}{cc}
\mathcal A_+|n\rangle=\sqrt{[n+1]}|n\rangle,\\
\mathcal A_-|n\rangle=\sqrt{[n]}|n-1\rangle,\\
\mathcal N|n\rangle=n|n\rangle.
\end{array}
\end{eqnarray}
Note that in the Fock space there exists a deforming map \cite{Ku} to the classical
$\mathcal U'(h_4)$ given by
\begin{eqnarray}\label{3-11}
\mathcal A_-=\sqrt{\frac{[N+1]}{N+1}}A_-,\,\,\mathcal A_+=A_+\sqrt{\frac{N+1}{N+1}},\,\,\mathcal
N=N.
\end{eqnarray}
The operator
\begin{eqnarray}\label{3-12}
{\mathcal H}_2=\frac{\omega}{2}\left(\mathcal A_+\mathcal A_-+\mathcal A_-\mathcal A_+\right)=\frac{\omega}{2}([\mathcal
N+1]+[\mathcal N]),
\end{eqnarray}
can be considered to be the $q$-analog of the scalar field part of super-Hamiltonian
defined in (\ref{2-2}). Furthermore, the phase space realization of the $q$-deformed
oscillator is given by
\begin{eqnarray}\label{3-13}
\begin{array}{cc}
\hat x_2=\frac{1}{\sqrt{2M\omega}}(\mathcal A_++\mathcal A_-),\\
\hat\Pi_2=i\sqrt{\frac{M\omega}{2}}(\mathcal A_+-\mathcal A_-).
\end{array}
\end{eqnarray}
Hence, the commutation relation
\begin{eqnarray}\label{3-131}
[\hat x_2,\hat\Pi_2]=i([\mathcal N+1]-[\mathcal N]),
\end{eqnarray}
shows that the effective Planck's constant is no longer a constant, but depends
on the state of the scalar field.

Let us now  consider the case in which $q$ is a primitive root of unity, i.e.,
\begin{eqnarray}\label{3-14}
q=\exp\left(\frac{2\pi i}{\mathfrak{N}}\right),
\end{eqnarray}
where $\mathfrak{N}$ is a natural number, $\mathfrak{N}\in\mathbb{N}^+$, and $\mathfrak{N}\geq2$. {
 In the reminder of this section
we will show that it is equivalent to the existence of  the finite number of possible quantum states in the
Universe.

It is clear that for  $\mathfrak{N}\rightarrow\infty$, the deformation parameter
$q\rightarrow1$ and all of the deformed quantities will reduce to the ordinary
undeformed  ones.} The
quantum number defined in (\ref{3-61}) will be
\begin{eqnarray}\label{3-15}
[y]=\frac{q^{\frac{1}{2}y}-q^{-\frac{1}{2}y}}{q^\frac{1}{2}-q^{-\frac{1}{2}}}=\frac{\sin\left(\frac{\pi y}{\mathfrak{N}}\right)}{\sin\left(\frac{\pi}{\mathfrak{N}}\right)}.
\end{eqnarray}
Evidently, $q^\mathfrak{N}=1$, $[\mathfrak{N}]=[k\mathfrak{N}]=0$ and $[\mathfrak{N}+k]=[k]$ where $k$ is an integer.
By using identities (\ref{3-9}) and the defining relations of $\mathcal U'_q(h_4)$
on can show that at the root of unity the elements $\{\mathcal
A_+^\mathfrak{N}, \mathcal A_-^\mathfrak{N}, q^{\frac{\mathfrak{N}}{2}\mathcal N}, q^{-\frac{\mathfrak{N}}{2}\mathcal N}\}$ lie in the center of $\mathcal U'_q(h_4)$. The action of pairs of operators
$\{\mathcal A_+, \mathcal A_-\}$ on the basis eigenvectors are
\begin{eqnarray}\label{3-16}
\begin{array}{cc}
\mathcal A_+|n\rangle=\sqrt{[n+1]}|n\rangle,\\
\mathcal A_-|n\rangle=\sqrt{[n]}|n-1\rangle,\\
\mathcal A_-|0\rangle=0,\,\,\,\,\,\,\mathcal A_+|\mathfrak{N}-1\rangle=0.
\end{array}
\end{eqnarray}
Therefore, $\mathcal A_+$  annihilate the state $|\mathfrak{N}-1\rangle$ and the Fock space $\mathcal F_2(q)$ is finite $\mathfrak N$-dimensional vector
space with basis $\{|0\rangle,|1\rangle,...,|\mathfrak{N}-1\rangle\}$.
The Fock space matrix representation of the generators then become finite
dimensional \cite{Bon}
\begin{eqnarray}\label{3-17}
\begin{array}{cc}
\mathcal A_+=\sum_{n=0}^{\mathfrak{N}-2}\sqrt{[n+1]}|n+1\rangle\langle n|,\\
\\
\mathcal A_-=\sum_{n=1}^{\mathfrak{N}-1}\sqrt{[n]}|n-1\rangle\langle n|,\\
\\
\mathcal N=\sum_{n=0}^{\mathfrak{N}-1}n|n\rangle\langle n|.
\end{array}
\end{eqnarray}
This representation follows the nilpotency of the generators of $\mathcal
U_q'(h_4)$
\begin{eqnarray}\label{3-18}
\mathcal A^\mathfrak{N}_{\pm}=0.
\end{eqnarray}
Now, Eqs.(\ref{3-12}) and (\ref{3-15}) admit the following eigenvalues for $\mathcal
H_2$
\begin{eqnarray}\label{3-19}
E_n=\frac{\omega}{2}\frac{\sin(\frac{\pi}{\mathfrak{N}}(n+\frac{1}{2}))}{\sin(\frac{\pi}{2\mathfrak{N}})},\,\,n=0,...,\mathfrak{N}-1.
\end{eqnarray}
Note that for $\mathfrak{N}\rightarrow\infty$ the earlier eigenvalues will reduce to
(\ref{2-3}). Since $\sin(\frac{\pi}{\mathfrak{N}}(n+\frac{1}{2}))=\sin(\frac{\pi}{\mathfrak{N}}(\mathfrak{N}-n-1+\frac{1}{2}))$,
there is a two-fold degeneracy at the eigenvalues.

An interesting question is the determination of the
eigenvalues of the pairs operators
$\{\hat x_2,\hat\Pi_2\}$ and the corresponding eigenvalues. We shall perform
this for the scalar field operator $\hat x_2$ (for corresponding momenta the analysis is similar). Let $|x_2\rangle$ and $x_2$ be the eigenvector
and the corresponding eigenvalue of the operator $\hat x_2$, satisfying
\begin{eqnarray}\label{3-20}
\hat x_2|x_2\rangle=x_2|x_2\rangle.
\end{eqnarray}
As we mentioned before, the Fock space is a $\mathfrak N$-dimensional $\mathbb{C}(q)$-vector
space. Therefore, the expression of $|x_2\rangle$ in the $q^{\mathcal N}$  representation
will be
\begin{eqnarray}\label{3-21}
|x_2\rangle=\sum_{n=0}^{\mathfrak{N}-1}\frac{c_n(x_2)}{\sqrt{2^n[n]!}}|n\rangle.
\end{eqnarray}
On the other hand, Eqs.(\ref{3-13}) and (\ref{3-16}) give
\begin{eqnarray}\label{3-22}
\begin{array}{cc}
\hat x_2|n\rangle=\\\frac{1}{\sqrt{2M_\text{P}\omega}}\left(\sqrt{[n]}|n-1\rangle+\sqrt{[n+1]}|n+1\rangle\right).
\end{array}
\end{eqnarray}
By Eqs. (\ref{3-20})-(\ref{3-22}) one obtains the
following recurrence relations for the coefficients $c_n$
\begin{eqnarray}\label{47}
\begin{array}{cc}
c_1(x_2)=2\sqrt{M_\text{P}\omega}x_2c_0(x_2),\\
c_{n+1}(x_2) =2\sqrt{M_\text{P}\omega}xc_n(x_2) -2 [n]c_{n-1}(x_2),\\
c_\mathfrak{N}(x_2)=0.
\end{array}
\end{eqnarray}
These relations have the solutions \cite{Bon}
\begin{eqnarray}\label{48}
c_n(x_2)  =H_n(\sqrt{M_\text{P}\omega}x_2;q^\frac{1}{2}),
\end{eqnarray}
where $H_n(y;q^\frac{1}{2})$ are the $q$-Hermite polynomials satisfy the equation
\begin{eqnarray}\label{49}
\begin{array}{cc}
H_{n+1}(y;q^\frac{1}{2})+\\+2[n]H_{n-1}(y;q^\frac{1}{2})-2yH_n(y;q^\frac{1}{2})=0,
\end{array}
\end{eqnarray}
The last condition in (\ref{47}) is equivalent to
\begin{eqnarray}\label{50}
H_\mathfrak{N}(\sqrt{M_{P}\omega}x_{2,\mu};q^\frac{1}{2})=0.
\end{eqnarray}
where $x_{2,\mu}$ are the roots of $q$-Hermite
polynomial. Hence, the eigenvalues of operator $\hat x_2$ in a $\mathfrak{N}$-dimensional Fock space  are the roots of the corresponding $q$-Hermite polynomial.
Also, the $q$-Hermite polynomials defined by the recurrence
relation of Eq.(\ref{49}) are odd (or even) functions of $x_2$
for n odd (or even), therefore the half of eigenvalues of $x$ are
negative.
 The number of these
roots is equal to the order $\mathfrak{N}$ of the corresponding
polynomial and these roots are real \cite{Bon}. The discrete values of $q$-Hermite polynomial are indexed by the following convention
\begin{eqnarray}\label{52}
\mu=-l,-l+1,....,l-1,l,
\end{eqnarray}
where $\mathfrak{N}=2l+1$ for odd values of $\mathfrak{N}$ and $\mathfrak{N}=2l$ for even values of $\mathfrak{N}$.

The $q$-deformed Hermite polynomials for the first few values of n
are listed below:
\begin{eqnarray}\label{Hermite}
\begin{array}{cc}
H_1(y;q^\frac{1}{2})=2y,\\
H_2(y;q^\frac{1}{2})=4y^2-2,\\
H_3(y;q^\frac{1}{2})=8y^3-4[2][\frac{3}{2}]y,\\
H_4(y;q^\frac{1}{2})=16y^4-8[2]^2[\frac{3}{2}]y^2+4[3],\\
H_5(y;q^\frac{1}{2})=\\=32y^5-16[2]^2[\frac{5}{2}]y^3+8([3]+[2][\frac{3}{2}][4])y.
\end{array}
\end{eqnarray}
In simplifying the above relations we used $[1]+[2]+...+[n]=[2][\frac{n+1}{2}][\frac{n}{2}]$.
With similar analysis, one can show that $q$ being a root of unity induces
a discretization of the spectrum of the momenta
\begin{eqnarray}\label{50b}
H_\mathfrak{N}\left(\frac{1}{\sqrt{M_\text{P}\omega}}\Pi_{2,\mu};q^\frac{1}{2}\right)=0,
\end{eqnarray}
which shows that the spectra of the operators $\sqrt{M_\text{P}\omega}\hat x_2$ and $\frac{1}{M_\text{P}\omega}\hat\Pi_2$ being identical and the same q-Hermite polynomials appearing
in both cases.

As a result of the discretization of the ``position"
and ``momenta" eigenvalues found above, the phase
space of scalar field, $(\hat x_2,\hat\Pi_2)$, is not the whole real plane, but it is a two-dimensional
lattice with non-uniformly distributed
points. For very large values of $\mathfrak{N}$ the $q$-numbers are reduced to the ordinary reals, so $[\mathfrak{N}]\simeq \mathfrak{N}$ and consequently the $q$-Hermite and defined in Eqs.(\ref{49}) will be reduced to the ordinary Hermite polynomials. Also, the largest root of Hermite
polynomial $H_\mathfrak{N}(y)$ is $y_{\mathfrak{N},\mathfrak{N}}\simeq\sqrt{2\mathfrak{N}}$ \cite{Her}. Therefore, according
to Eq.(\ref{50}) for the large
values of $\mathfrak{N}$ the largest value of scalar field is
$x_\text{2max}\simeq L_\text{P} \sqrt{\mathfrak{N}}$.

\subsection{Quantum deformation of the gravitational sector}

Let us now investigate the quantum deformation of the gravitational part of super-Hamiltonian,
$\mathcal H_1$ defined in Eq.(\ref{2-4}).
We should write the Heisenberg-Weyl Lie algebra for gravitational part with generators
$\{\bar A_+,\bar A_-,\bar N\}$
\begin{eqnarray}\label{4-1}
[\bar A_-,\bar A_+]=1,\,\,[\bar N,\bar A_{\pm}]=\pm\bar A_{\pm}
\end{eqnarray}
In the Fock space, $\mathcal F_1$, with the basis $\{|n\rangle,\,\,\bar N|n\rangle=n|n\rangle\}$
the pairs of operators $\bar A_\pm$ act due
\begin{eqnarray}\label{3-4b}
\begin{array}{cc}
\bar A_+|n\rangle=\sqrt{n+1}|n+1\rangle,\\
\bar A_-|n\rangle=\sqrt{n}|n-1\rangle.
\end{array}
\end{eqnarray}
But, as we showed is section II, the gravitational part of super-Hamiltonian
${\mathcal H}_1$ has the self-adjoint extension if the wave function obeys the standard Dirichlet or Neumann boundary conditions.

{  The states of  the Fock space are thus classified into two disjoint odd and even subspaces. Therefore,
it seems that the $h_4$ does not represent the suitable symmetry of the gravitational
part.} To split the Hilbert space into odd and even subspaces { and obtain
the true symmetry of gravitational part,} let us introduce
the generators
\begin{eqnarray}\label{4-5}
K_0=\frac{1}{2}(\bar N+\frac{1}{2}),\,\,\,\,\,\,\,\,\,K_\pm=\frac{1}{2}\left(\bar A_\pm\right)^2.
\end{eqnarray}
It is not difficult to verify that these generators satisfy the  commutation relations
\begin{eqnarray}\label{5-6}
[K_0,K_\pm]=\pm K_\pm,\,\,\,\,\,\,\,\,\,\,[K_+,K_-]=-2K_0,
\end{eqnarray}
of the algebra $su(1,1)$.
The $*$-involution  on the elements of algebra is defined by $K_0^*=K_0$ and
$K_+^*=-K_-.$ The positive
discrete series representations of this Lie algebra, $D_k^+$, are
labeled by a positive real number $k>0$. Suppose that $\{|k,m\rangle, m = 0, 1, 2, ...\}$ is the basis in
the Hilbert space $V_k$ of representation $D_k^+$.  Then, the actions of the above generators on a set of basis eigenvectors $|k,m\rangle$ are given by
\begin{equation}\label{5-7}
\begin{array}{cc}
K_0|k,m\rangle=(k+m)|k,m\rangle,\\
K_+|k,m\rangle=\sqrt{(2k+m)(m+1)}|k,m+1\rangle,\\
K_-|k,m\rangle=\sqrt{(2k+m-1)m}|k,m-1\rangle.
\end{array}
\end{equation}
The Casimir operator, ${\mathcal C}_2$, is a central
self-adjoint element of the universal enveloping algebra
${\mathcal U'}(su(1,1))$
and
\begin{equation}\label{5-8}
\begin{array}{cc}
\mathcal C_2:=K_0(K_0+1)-K_-K_+,\\
\mathcal C_2|k,m\rangle=k(k-1)|k,m\rangle.
\end{array}
\end{equation}
In addition, the gravitational part of super-Hamiltonian
(\ref{2-4}) can be presented as
\begin{eqnarray}\label{5-81}
\mathcal H_1=2\omega K_0,
\end{eqnarray}
which leads us to point out that the Casimir operator
commutes with $\mathcal H_1$. A direct calculation of the Casimir operator (\ref{5-8}) with the aid of (\ref{4-5}) shows that the
eigenvalue $k(k-1)$ in this particular case is equal to $-\frac{3}{16}$. This means that  $k$ is equal to either $k=\frac{1}{4}$ or $k=\frac{3}{4}$ where each of these two values of $k$ defines a unitary
irreducible representation of the algebra $su(1,1)$: $D_\frac{1}{4}^+$ consists of those eigenstates of $\mathcal H_1$, which correspond to the eigenvalues $k_1 +n = n+\frac{1}{4} = \frac{1}{2}(2n+\frac{1}{2})$,
$n = 0, 1, 2, ...,$ of the generator $\mathcal K_0$; whereas $D^+_{\frac{3}{4}}$ corresponds to the
eigenvalues $k_2 +n = n+\frac{3}{4}=\frac{1}{2} (2n+1+\frac{1}{2})$.

As we make progress, let us  summarize the q-deformation of the universal enveloping algebra of $su(1,1)$, $\mathcal U'_q(su(1,1))$, when the deformation parameter $q$ is generic. The quantized enveloping algebra
$\mathcal U'_q(su(1,1))$  is an associative unital algebra generated by the four generators $\{\mathcal K_+,\mathcal K_-,q^{\mathcal K_0},q^{-\mathcal K_0}\}$ with the commutation
relations
\begin{eqnarray}\label{5-9}
\begin{array}{cc}
[\mathcal K_+,\mathcal K_-]=-\frac{q^{2\mathcal K_0}-q^{-2\mathcal K_0}}{q-q^{-1}},\\

[\mathcal K_0,{\mathcal K}_\pm]=\pm{\mathcal K}_\pm.
\end{array}
\end{eqnarray}
These relations reduce to those of algebra $su(1,1)$ defined
in (\ref{5-6}) in the classical limit $q\rightarrow1$. $\mathcal U'_q(su(1,1))$
is equipped with a structure of Hopf algebra: $\Delta(\mathcal K_0)=\mathcal
K_0\otimes\mathcal K_0$, $\Delta(\mathcal K_\pm)=\mathcal K_\pm\otimes\mathcal
K_0+\mathcal K_0^{-1}\otimes\mathcal K_\pm$, $\epsilon(\mathcal K_0)=1$,
$\epsilon(\mathcal K_\pm)=0$, $S(\mathcal K_0)=\mathcal K_0^{-1}$ and $S(\mathcal
K_\pm)=-q^{\mp1}\mathcal K_\pm$.
Also the Casimir operator is
\begin{eqnarray}\label{5-10}
\mathcal C_2=[\mathcal K_0+\frac{1}{2}]_q^2-\mathcal K_-\mathcal K_+=[\mathcal K_0+\frac{1}{2}]_q^2+\mathcal K_+\mathcal K_-,
\end{eqnarray}
where
\begin{eqnarray}\label{5-11}
[y]_q=\frac{q^y-q^{-y}}{q-q^{-1}}.
\end{eqnarray}
Suppose that $\{|k, m\rangle, m = 0, 1, 2, ...\}$ is the basis in
the Hilbert space $\mathcal F_1(k)$ of representation $D^+_k$. The action of the generators then has the form \cite{Dama}
\begin{eqnarray}\label{5-12}
\begin{array}{cc}
\mathcal K_0|k,m\rangle=(m+k)|k,m\rangle,\\
\mathcal K_+|k,m\rangle=\sqrt{[m+1]_q[m+2k]_q}|k,m+1\rangle,\\
\mathcal K_-|k,m\rangle=\sqrt{[m]_q[m+2k-1]_q}|k,m-1\rangle.
\end{array}
\end{eqnarray}
Also
\begin{eqnarray}\label{5-13}
\begin{array}{cc}
\mathcal C_2|k,m\rangle=[k-\frac{1}{2}]_q^2|k,m\rangle,\\

[\mathcal K_0]_q|k,m\rangle=[k+m]_q|k,m\rangle.
\end{array}
\end{eqnarray}
The elements of this basis are obtained from the highest vector
$|k,0\rangle$ by a second application of the operator $\mathcal K_+$,
\begin{eqnarray}\label{5-14}
|k,m\rangle=\frac{1}{\sqrt{[m]_q!([2k]_q)_m}}\mathcal K_+^m|k,0\rangle,
\end{eqnarray}
where
\begin{eqnarray}\label{5-15}
([y]_q)_m=[y]_q[y+1]_q...[y+m-1]_q,
\end{eqnarray}
is the Pochgammer $q$-symbol.
The generators of $\mathcal U_q '(su(1,1))$ can be realized \cite{Ku} with the aid of the generators of the algebra $\mathcal U_q'(h_4)$
\begin{eqnarray}\label{5-161}
\mathcal K_0=\frac{1}{2}(\bar{\mathcal N}+\frac{1}{2}),\,\,\,\,\,\,\mathcal K_\pm=\frac{1}{[2]}\left(\bar{\mathcal
A}_\pm\right)^2.
\end{eqnarray}
In this case the Fock space representation of the $q$-oscillator splits into
the direct sum of two irreducible components $\mathcal F_1=\mathcal F_e\oplus\mathcal
F_o$, where $\mathcal F_e(\mathcal F_o)$ is formed by states
with an even (odd) number of quanta. Using (\ref{5-161}) and (\ref{5-10}) we obtain
\begin{eqnarray}\label{5-17}
\mathcal C_2|n\rangle=[\frac{1}{4}]^2_q|n\rangle=[-\frac{1}{4}]_q^2|n\rangle,
\end{eqnarray}
where $|n\rangle\in\mathcal F_1$. It then follows from (\ref{5-10}) that $k =\frac{1}{4}$ or $k =\frac{3}{4}$. Taking
(\ref{5-12}) into account, we see that the representation with $k =\frac{1}{4}$ and $|2m\rangle = |\frac{1}{4}, m\rangle$ acts in
$\mathcal F_o$ and in $\mathcal F_e$ we have $k =\frac{3}{4}$ and $|2m+1\rangle = |\frac{3}{4}, m\rangle$ like as the classical $\mathcal U'(su(1,1))$.

Let us now return to the case in which $q$ is a primitive root of unity as
defined
in (\ref{3-14}). The
new quantum number defined in Eq.(\ref{5-11}) will be
\begin{eqnarray}\label{3-151}
[y]_q=\frac{q^{y}-q^{-y}}{q-q^{-1}}=\frac{\sin\left(\frac{2\pi y}{\mathfrak{N}}\right)}{\sin\left(\frac{2\pi}{\mathfrak{N}}\right)}.
\end{eqnarray}
If we define, $l$ as:
 $\mathfrak{N}=2l+1$ for odd values of $\mathfrak N$ and $\mathfrak{N}=2l$ for even values of $\mathfrak{N}$, then for
 odd values of $\mathfrak N$ and $m=l-1$ ($k=\frac{3}{4}$) we obtain $[m+2k]_q=[\frac{\mathfrak{N}}{2}]_q=0$. Also
 for even values of $\mathfrak{N}$ and $m=l$ ($k=\frac{1}{4}$) we have $[m]=[l]_q=[\frac{\mathfrak{N}}{2}]_q=0$. Therefore,
 elements $\{\mathcal K_+^l,\mathcal K_-^l,q^{\pm l\mathcal K_0}\}$ lie in the
 center of $\mathcal U'_q(su(1,1))$. These
 lead us to
 \begin{eqnarray}\label{5-16}
 \mathcal K_+|k,l-1\rangle=0.
 \end{eqnarray}
Hence, the Fock space split into two $l$-dimensional spaces with bases $\{|\frac{1}{4},m\rangle, m=0,...,l-1\}$ and $\{|\frac{3}{4},m\rangle, m=1,...,l\}$.

The
domain of definition of scale factor, $x_1 = a$, is $\mathbb{R}^+$, so the
conjugate momenta $\Pi_1$ is not Hermitian, but one can easily show that
$\Pi^2_1$ is Hermitian. Let us then obtain the eigenvalues and eigenvalues of $\hat x^2_1$ and $\hat
\Pi^2_1$. Let $|x_1\rangle$ ($|\Pi_1\rangle$) and $x^2_1$ ($\Pi^2_1$) be the eigenvector
and eigenvalue of the operator $\hat x^2_1$ ($\hat \Pi^2_1$), satisfying
\begin{eqnarray}\label{5-17a}
\begin{array}{cc}
\hat x^2_1|x_1\rangle=x_1^2|x_1\rangle,\\
\hat \Pi^2_1|\Pi_1\rangle=\Pi_1^2|\Pi_1\rangle.
\end{array}
\end{eqnarray}
Likewise  to the scalar field part, the expression of $|x_1\rangle$ ($|\Pi_1\rangle$) in the $\mathcal K_0$ representation is given by
\begin{eqnarray}\label{5-18}
\begin{array}{cc}
|x_1\rangle=\sum_nc'_n (-1)^n\sqrt{\frac{[n]_q!}{[n+2k-1]_q!}}|k,n\rangle,\\
|\Pi_1\rangle=\sum_nc''_n (-1)^n\sqrt{\frac{[n]_q!}{[n+2k-1]_q!}}|k,n\rangle,
\end{array}
\end{eqnarray}
where the $q$-factorial defined by $[n]_q!=\prod_{m=1}^n[m]_q$.
Using the phase space realizations
\begin{eqnarray}\label{5-19}
\begin{array}{cc}
\hat x_1=\frac{1}{\sqrt{2M_\text{P}\omega}}({{\bar \mathcal A_+}}+{\bar\mathcal A_-}),\\
\hat\Pi_1=i\sqrt{\frac{M_\text{P}\omega}{2}}({\bar \mathcal A_+}-{\bar\mathcal A_-}),
\end{array}
\end{eqnarray}
and (\ref{5-161}) we obtain
\begin{eqnarray}\label{5-20}
\begin{array}{cc}
\frac{2}{M_\text{P}\omega}\hat\Pi_1^2=\frac{1}{[\frac{1}{2}]_q}(\mathcal K_++\mathcal
K_-)-\frac{[\frac{1}{2}]_q}{[\frac{1}{4}]_q}[\mathcal K_0]_q,\\
2M_\text{P}\omega\hat x_1^2=\frac{1}{[\frac{1}{2}]_q}(\mathcal K_++\mathcal
K_-)+\frac{[\frac{1}{2}]_q}{[\frac{1}{4}]_q}[\mathcal K_0]_q.
\end{array}
\end{eqnarray}
Eqs. (\ref{5-12}), (\ref{5-17}), (\ref{5-18}) and (\ref{5-20}) give recursion
relations
\begin{eqnarray}\label{5-21}
\begin{array}{cc}
\frac{1}{2[\frac{1}{2}]_q}[n+1]_qc'_{n+1}=\left(\frac{[\frac{1}{2}]_q}{2[\frac{1}{4}]_q}[k+n]_q-M\omega
x_1^2\right)c_n'\\
-\frac{1}{2[\frac{1}{2}]_q}[n+2k-1]_qc'_{n-1},\\
c'_0=1,\\
c'_l=0,
\end{array}
\end{eqnarray}
which is the  recursion relation for the $q$-deformed generalized Laguerre polynomials
\begin{eqnarray}\label{5-23}
\begin{array}{cc}
c_n'=L_n^{(2k-1)}(M_\text{P}\omega x_1^2;q),\\
L_l^{(2k-1)}(M_\text{P}\omega x_1^2;q)=0.
\end{array}
\end{eqnarray}
 Note that for
$q=1$ it reduce to the recursion relation of the ordinary generalized Laguerre polynomials. Hence, the eigenvalues of square of scale factor $\hat x^2_2$ and $\Pi_1^2$ in a $l$-dimensional Fock space  are the roots of the corresponding generalized $q$-Laguerre polynomials
\begin{eqnarray}\label{mama}
\begin{array}{cc}
L_l^{(2k-1)}(M_\text{P}\omega x^2_{1,\mu};q)=0,\\
L_l^{(2k-1)}(\frac{\Pi^2_{1,\mu}}{{M_\text{P}\omega}};q)=0,
\end{array}
\end{eqnarray}
where $x^2_{1,\mu}$ and $\Pi^2_{1,\mu}$ are the positive roots of $q$-Laguerre
polynomial.
One can show that the $q$-deformed Hermite polynomials defined in recursion
relation (\ref{49}) satisfy following relations

\begin{equation}\label{5-24}
\begin{array}{cc}
H_{2n+2}(y;q^\frac{1}{2})=\\4\left(y^2-\frac{1}{2[\frac{1}{4}]_q}[n+\frac{1}{4}]_q\right)H_{2n}(y;q^\frac{1}{2})\\-\frac{4}{[\frac{1}{2}]^2_q}[n]_q[n-\frac{1}{2}]_qH_{2n-2}(y;q^\frac{1}{2}),\\
H_{2n+3}(y;q^\frac{1}{2})=\\4\left(y^2-\frac{1}{2[\frac{1}{4}]_q}[n+\frac{3}{4}]_q\right)H_{2n+1}(y;q^\frac{1}{2})\\-\frac{4}{[\frac{1}{2}]^2_q}[n]_q[n+\frac{1}{2}]_qH_{2n-1}(y;q^\frac{1}{2}).
\end{array}
\end{equation}
Using these recursion relations, it is easy to show that the generalized $q$-Laguerre polynomials are related to the $q$-Hermite polynomials
\begin{eqnarray}\label{5-25}
\begin{array}{cc}
H_{2n}(y;q^\frac{1}{2})=\\\left(\frac{2}{[\frac{1}{2}]_q}\right)^n(-1)^n[n]_q!L_n^{(-\frac{1}{2})}(y^2;q);\\
H_{2n+1}(y;q^\frac{1}{2})=\\y\left(\frac{2}{[\frac{1}{2}]_q}\right)^{n+\frac{1}{2}}(-1)^n[n]_q!L_n^{(\frac{1}{2})}(y^2;q).
\end{array}
\end{eqnarray}
These relations show that the eigenvalues of square of scale factor and conjugate momenta
are $l$ positive, non-zero roots of
$q$-Hermite polynomials for even and odd values of $\mathfrak{N}$ given by
\begin{equation}\label{2-26}
\begin{array}{cc}
H_\mathfrak{N}(y;q^\frac{1}{2})=0,\hspace{.3cm}\text{if $\mathfrak{N}$ is even},\\
\frac{H_\mathfrak{N}(y;q^\frac{1}{2})}{y}=0,\hspace{.4cm}\text{if $\mathfrak{N}$ is odd},
\end{array}
\end{equation}
where $y\in\{\sqrt{M_\text{P}\omega} x_{1,\mu},\frac{\Pi_{1,\mu}}{\sqrt{M_\text{P}\omega}}\}$.
The second equation shows that for odd values of $\mathfrak{N}$ the zero eigenvalue
is removed from the set of eigenvalues of scale factor and conjugate momenta.

In order to get  the discrete eigenvalue spectrum of the four operator $\{x^2_1,x_2,\Pi^2_1,\Pi_2\}$, let us consider the
special case $\mathfrak{N} =5$ as an example. Then the roots of $q$-Hermite polynomials
defined in Eqs.(\ref{50}) and (\ref{50b})  give us the following approximate eigenvalues for
$x_2$ and corresponding conjugate momenta
\begin{equation}\label{53}
\begin{array}{cc}
\frac{x_2}{L_\text{P}}\simeq\{-0.12,-0.07,0,0.07,0.12\},\\
L_\text{P}\Pi_2\simeq\{-13.94,-8.33,0,8.33,13.94\}.
\end{array}
\end{equation}

Also, the roots of  $q$-Laguerre polynomials (\ref{mama}), or equivalently the second equation in (\ref{2-26}),
give us the eigenvalues of scale factor and the square of the conjugate momenta
\begin{eqnarray}\label{last}
\begin{array}{cc}
\frac{x_1}{L_\text{P}}\simeq\{0.07,0.12\},\\
L^2_\text{P}\Pi_1^2\simeq\{69.37,194.41\}.
\end{array}
\end{eqnarray}
where  $L_\text{P}=\sqrt{8\pi G}$ is the reduced Planck length.

\section{The IR/UV mixing and ``new'' cosmological constant problem}

{ Let us briefly review the ``old'' and the ``new'' CC problems.
Regarding locality and unitarity of quantum field theory, the vacuum has an energy. To obtain the corresponding energy we need to calculate the vacuum loop diagrams for each matter field species. For example the one loop diagram result for a scalar field of mass $m$ is given by \cite{Martin}
\begin{eqnarray}\label{Q1}
V_{\text{vac}}^{\text{1-loop}}\simeq-\frac{m^4}{(8\pi)^2}\left(\frac{2}{\epsilon}+\ln\left(\frac{M_\text{UV}^2}{m^2}\right)+\text{finite}\right),
\end{eqnarray}
where we work in $d=4-\epsilon$ dimensions and $M_\text{UV}$ is the UV regulator scale. If we add a counter term
\begin{eqnarray}\label{Q2}
V_{\text{c}}^{\text{1-loop}}\simeq\frac{m^4}{(8\pi)^2}\left(\frac{2}{\epsilon}+\ln\left(\frac{M_\text{UV}^2}{M^2}\right)+\text{finite}\right),
\end{eqnarray}
where $M$ is the renormalisation scale to eliminate the divergences, then the renormalised vacuum energy at 1-loop level will be
\begin{eqnarray}\label{Q3}
\Lambda_\text{ren}^\text{1-loop}\simeq\frac{m^4}{(8\pi)^2}\left(\ln\left(\frac{m^2}{M^2}\right)+\text{finite}\right).
\end{eqnarray}
It is clear that the finite contributions to vacuum energy are completely arbitrary since they can always
be absorbed into a re-definition of the subtraction scale $M$. This emphasises
that QFT does not have a concrete prediction for the renormalised vacuum energy. In comparison to $V_\text{vac}^{1-loop}$, the counter term $\Lambda_\text{c}^\text{1-loop}$
 has a divergent and finite part where this finite part can be consider as
 the ``bare'' classical CC that we were free to add to
the Einstein-Hilbert action. Since QFT cannot theoretically predict the magnitude  of the CC, we have to
measure it and adjust the finite part of $\Lambda_\text{c}$ appropriately such that the theory
matches with cosmological observations. For example, if we assume that our scalar field is the
Higgs boson of the standard model then it has a mass $m = 126$GeV. Given that observations place an upper bound on the total CC of $(\text{meV})^4\simeq10^{-60}(\text{TeV})^4$, the finite contributions to the vacuum energy at the 1-loop
level must cancel to an accuracy of 1 part in $10^{60}$. Let us now consider the 2-loop correction to the vacuum energy from the
massive scalar field. At two loops, we consider the so-called scalar ``figure of
eight'' with external graviton legs. Its contribution to vacuum energy is given by
\begin{eqnarray}\label{Q4}
V_\text{vac}^{2-loop}\simeq\lambda m^4.
\end{eqnarray}
For perturbative theories without finely tuned couplings, where $\lambda\sim\mathcal O(0.1)$, (as for
example the Standard Model Higgs) this is a
huge contribution to the cosmological constant relative to the observed value. Now having already fixed the bare CC to match
observations at the 1-loop level, we need to re-tune its value to a high order of
accuracy to cancel the unwanted contributions at the 2-loop level. Similarly
we have to re-tune its value as we go to 3-loops and so on.  In other words, it is radiatively
unstable and we need to re-tune the bare CC at every order in loop perturbation theory to deal with this instability.

 Because this radiatively instability of vacuum energy was already a problem before the late time acceleration of Universe  was discovered,
this is sometimes called the ``old'' CC problem \cite{Wi}. In fact, it is distinguished from
a ``new'' problem that is more to do with
understanding the precise value and the origin of the CC that has been observed. The ``new'' CC problem \cite{Bur} has its origin in the discovery that the vacuum energy is not exactly zero and it causes observed late time acceleration of Universe. In this direction,  it is usually presupposed that there is a solution to the ``old''
CC problem that makes the vacuum energy precisely zero and radiatively stable \cite{Li}. Then it would be remain to
 explain why the measured CC is not precisely zero, and instead has a nonzero but very small value. The ``new'' version of the CC problem divides itself into two parts: $i$) why it is so small? $ii$) How come that the energy density of CC and baryonic matter have the same order of magnitude at the present epoch? This is called the coincidence problem. Let us now describe how the ``new''
CC problem can be discussed from quantum deformation of quantum cosmology perspective \cite{J}. }

 For very large values of $\mathfrak{N}$ the $q$-numbers are reduced to the ordinary reals, so $[\mathfrak{N}]\simeq[\mathfrak{N}]_q\simeq \mathfrak{N}$ and consequently the $q$-Hermite and  $q$-Laguerre
polynomials defined in Eqs.(\ref{49}) and (\ref{5-21}) will be reduced to the ordinary Hermite and Laguerre polynomials.
{Let $\chi_{n,k}^{(\alpha)}$, $k =1,2,...$, denote the zeros of the Laguerre polynomial $L^{(\alpha)}_n(y)$, in increasing order, for large values of
$n$.  It is well-known that these zeros lie in the oscillatory region $0<\chi_{n,k}^{(\alpha)}<4n+2\alpha+2$. Then the smallest root, $\chi^{(\alpha)}_{n,1}$ and the largest, $\chi^{(\alpha)}_{n,n}$,  are given by \cite{La}
\begin{eqnarray}\label{lar}
\chi^{(\alpha)}_{n,1}\simeq\frac{(\alpha+1)(\alpha+3)}{2n+\alpha+1},\,\,\,\,\,\,\,\chi^{(\alpha)}_{n,n}\simeq4n.
\end{eqnarray}

Using the above  results in Eqs.(\ref{mama}) it is easy to show the minimum and
maximum eigenvalues of the square of the scale factor and the corresponding momenta
\begin{eqnarray}\label{maxima}
\begin{array}{cc}
a^2_\text{min}\simeq\frac{L^2_\text{P}}{\mathfrak{N}},\,\,\,\,\,\,\,\,\,\,\,\,\,\,\,\,\,a^2_\text{max}\simeq\mathfrak{N}L_\text{P}^2,\\
H^2_\text{min}\simeq\frac{1}{\mathfrak{N}L_\text{P}^2},\,\,\,\,\,\,\,\,\,\,H^2_\text{max}\simeq\frac{\mathfrak{N}}{L_\text{P}^2}.
\end{array}
\end{eqnarray}
The above relations thus suggest to introduce the smallest and largest distances respectively
as
\begin{eqnarray}\label{dis}
L_\text{min}=\frac{L_\text{P}}{\sqrt{\mathfrak N}},\,\,\,\,\,L_\text{max}=\sqrt{\mathfrak
N}L_\text{P}.
\end{eqnarray} }
These further suggest that the deformation parameter
(\ref{3-14}) may be rewritten as
\begin{eqnarray}\label{qq}
q=\exp{i\left(\frac{L_\text{P}}{L_\text{max}}\right)^2}.
\end{eqnarray}
 This yields a simple geometrical interpretation for the relation between the quantum deformation parameter and the maximal  possible value of the scale factor. In GR, a maximal distance of the order $L_\text{max}\simeq\Lambda^{-\frac{1}{2}}$ is also essentially implied  when a cosmological
constant $\Lambda$ is present. Hence, it seems that the quantum deformation
of cosmological model induce a cosmological constant and Eq.(\ref{qq}) should
be
\begin{eqnarray}\label{qq2}
q=\exp(i\Lambda L^2_\text{P}).
\end{eqnarray}

We can take the limit $q\rightarrow1$ by taking $\Lambda\rightarrow0$ which
leads to the original quantum cosmology without
cosmological constant\footnote{Loop quantum gravity (LQG) and spinfoam frameworks use the  cosmological constant as a coupling constant just like the gravitational constant.
A $q$-deformation has been derived in LQG as a way to implement
the dynamics of the theory with cosmological constant
 \cite{Han,Loop} and the deformation parameter, $q$, then is given by $(\ref{qq2})$
 \cite{Du}}. Let us stress that in our original cosmological model the CC does not exist but the quantum deformation of the model introduced
a cosmological constant, where it is related to the
natural number $\mathfrak{N}\in\mathbb{N}^+$ by
 \begin{eqnarray}\label{Lambda}
 \Lambda\simeq\frac{1}{L_\text{max}^2}=\frac{1}{\mathfrak{N}L^2_\text{P}}.
 \end{eqnarray}

{ In such a line of reasoning a CC should be understood as a direct consequence
of the finite number of states in the Hilbert space
which itself is a result of $q$-deformation.   The minimum value of scalar
curvature is $R_{min}\simeq\frac{1}{\mathfrak{N}L_\text{P}^2}\simeq\Lambda$. It immediately leads to a suggestion for
the  explanation for the late time acceleration of the Universe: the Universe reaches the minimally possible
curvature and has to stay in this state.}

{Let us now consider an observer located at  $3D$ sphere of curvature radius  $a_{\mathfrak{N},\mathfrak{N}}=L_\text{max}$.  If one measures the apparent size of a small sphere of diameter $r$ located at large distance $L$, $(r\ll L)$, an observer will see it under an angular size $\delta\phi\simeq\frac{r}{L}$.
Consider this sphere located at horizon (largest distance),  she(he) will
never see \cite{J} the sphere under an angle smaller than $\simeq\frac{r}{L_\text{max}}\simeq
r{\sqrt{\Lambda}}$.  Therefore, it is natural to assume that
it is impossible to measure areas having angular size smaller than
\begin{eqnarray}\label{length}
\delta\phi_\text{min}\simeq\frac{r_\text{min}}{L_\text{max}}\simeq\frac{L_\text{min}}{L_\text{max}}=\frac{1}{\mathfrak{N}}.
\end{eqnarray}
In such a situation, at the presence of minimal length $L_\text{min}$ everything that the observer sees is captured, on the
local celestial $2$-sphere formed by the directions around
him, by spherical harmonics with $j = j_\text{max}$.
In quantum groups language, a $2$-sphere not resolved at small angles is a Podle\'s ``quantum sphere'', $S^2_q$ \cite{Sphere}.
If we denote the generators of algebra $\mathcal A_q$ for $S^2_q$ by $\{\hat
X_+,\hat X_-,\hat X_3,1\}$
then they satisfy the following commutation relations \cite{Mad}
\begin{eqnarray}\label{s}
\begin{array}{cc}
\hat X_+\hat X_--\hat X_-\hat X_++\lambda\hat X_3^2=\mu\hat X_3,\\
q\hat X_3\hat X_+-q^{-1}\hat X_+\hat X_3=\mu\hat X_+,\\
q\hat X_-\hat X_3-q^{-1}\hat X_3\hat X_-=\mu \hat X_-,\\
\hat X_3^2+q\hat X_-\hat X_++q^{-1}\hat X_+\hat X_-=L_\text{max}^2,
\end{array}
\end{eqnarray}
where
\begin{eqnarray}\label{q}
\lambda=q-q^{-1},\,\,\,\,\,
\mu=L_\text{max}\frac{[2(\mathfrak{N}+1)]_q}{[\mathfrak{N}+1]_q\sqrt{[\mathfrak{N}]_q[\mathfrak{N}+2]_q}}.
\end{eqnarray}
  The above relations define quantum sphere $S^2_q$ when $q$ is root of unity.

 An interesting
limit of the above quantum sphere appears when $\mathfrak{N}$ is a very big natural
number. In
this case Eqs.(\ref{s}) will be reduce to
\begin{eqnarray}\label{fuzzy}
[\hat X_i,\hat X_j]=i\lambda_\mathfrak{N}\epsilon_{ij}^{\,\,\,\,k}\hat
X_k,\,\,\,\,\,\,\,\hat X_1^2+\hat X_2^2+\hat X_3^2=L_\text{max}^2,
\end{eqnarray}
where, $i,j,k=1,2,3$,
\begin{eqnarray}\label{lanbda}
\lambda_\mathfrak{N}=\frac{2L_\text{max}}{\sqrt{\mathfrak{N}(\mathfrak{N}+2)}}\simeq2L_\text{min},
\end{eqnarray}
and
\begin{eqnarray}\label{x}
\hat X_1=-\frac{1}{\sqrt2}(\hat X_++\hat X_-),\,\,\,\hat X_2=-\frac{i}{\sqrt2}(\hat
X_+-\hat X_-).
\end{eqnarray}
This algebra represents the fuzzy sphere $S^2_F$ \cite{F} which
can appear as vacuum solutions in Euclidean gravity \cite{Euc}.
The parameter $\lambda_\mathfrak{N}$ has dimension of length, and plays a role analogue to the
Planck's constant in quantum mechanics, as a quantization parameter. 
 From (\ref{fuzzy}) and definition of $\lambda_\mathfrak{N}$ and $L_\text{min}$ we can see that in the limit $\lambda_\mathfrak{N}\rightarrow0$
$(\mathfrak{N}\rightarrow\infty)$ the matrices $\hat X_i$ become commutative,
and we recover the  commutative sphere $S^2$ with radius $L_\text{max}\rightarrow\infty$.
In other words, at the limit $q\rightarrow1$, the emerged cosmological constant
and  corresponding noncommutative horizon will be disappear. Eqs.(\ref{s})
or alternatively Eqs.(\ref{fuzzy}) bear an interesting statement of holography.  In order to see the holography we argue in what follows that the smallest area that one can probe on $S^2_F$ is given by
\cite{Sheykh}
\begin{eqnarray}\label{area}
 Area(\text{min})=L_\text{min}L_\text{max}=L^2_\text{P}.
\end{eqnarray}
Furthermore, the surface area of $S^2_F$ is given by
\begin{eqnarray}\label{sur}
Area(S_F^2)=4\pi L_\text{max}^2\frac{\mathfrak{N}+1}{\sqrt{\mathfrak{N}(\mathfrak{N}+1)}}\simeq{4\pi L_\text{max}^2}.
\end{eqnarray}
Therefore, the number of smallest cells one can fit into the fuzzy sphere
of surface $L^2_\text{max}\simeq\frac{1}{\Lambda}$ is given by
\begin{eqnarray}\label{cell}
\frac{Area(S^2_F)}{Area(\text{min})}={\mathfrak{N}}.
\end{eqnarray}
This shows that the number of fundamental cell of the surface of fuzzy horizon
is equals to the dimension of Hilbert space of the scalar field.
The holographic principle asserts that the total number of degrees of freedom, or entropy $S_\text{dS}$, living on the holographic screen is bounded by one quarter of the area in Planck units
\begin{eqnarray}\label{entropy}
S_\text{dS}\simeq\frac{1}{\Lambda L_\text{P}^2}.
\end{eqnarray}
Using Eqs.(\ref{Lambda}), (\ref{s}) and (\ref{entropy}) one can readily check
that
\begin{eqnarray}\label{Hol}
S_\text{dS}\simeq ( \text{the number of cells
on the } S^2_F)\simeq \mathfrak{N},
\end{eqnarray}
which is basically the statement of holography. In this sense the
cosmological constant and holographic principle are emerged as the result
of the quantum deformation.

Moreover, noting that the minimum area defined in (\ref{area}) involves
both the UV character $L_\text{min}$ and the IR character $L_\text{max}$ one expects
the IR/UV mixing phenomenon.
It is generally assumed that particle physics can be accurately described by an EFT with an
ultraviolet UV cutoff, $M_\text{UV}$, less than the Planck mass, provided that all momenta and field strengths are small
compared with this cutoff to the appropriate power. Consequently the length $L$, which acts as an IR cutoff,
cannot be chosen independently of the UV cutoff, and scales relations obtained
in  Eq.(\ref{dis}). If $\rho_\Lambda\simeq M_\text{UV}^4$ is the quantum zero point energy density caused by a UV cutoff, the total energy of vacuum in a region of size $L$ should not exceed the maximum energy scale $\frac{1}{L_\text{min}}$, thus
\begin{eqnarray}\label{rho}
M_\text{UV}^4 L^3\leq\frac{1}{L_\text{min}}.
\end{eqnarray}
The largest $L=L_\text{max} $ allowed is the one saturating this
inequality. Thus
\begin{eqnarray}\label{last1}
 M_\text{UV}^4=\frac{1}{L_\text{min}L_\text{max}^3}=\mathfrak{N}^{-1}M_\text{P}^4.
\end{eqnarray}

{ To estimate the numerical value of the emerged CC obtained in (\ref{Lambda}) we need to know the value of the inverse of deformation parameter, $\mathfrak{N}$, which is equal to the entropy of degrees of freedom living on the emerged holographic screen. It is known \cite{Holo} that the total entropy of dust, $S_\text{(dust)}\simeq10^{80}$ and radiation $S_\text{(radiation)}\simeq10^{89}$ in observable Universe \cite{Chas} are related to the entropy of holographic screen, $S_\text{dS}$  via $S_\text{dS}\simeq S_\text{(dust)}^\frac{3}{2}\simeq S_\text{(radiation)}^\frac{4}{3}$. Hence, Eq.(\ref{Hol}) leads us to
\begin{eqnarray}\label{H111}
\mathfrak{N}=S_\text{dS}\simeq S_\text{(dust)}^\frac{3}{2}\simeq S_\text{(radiation)}^\frac{4}{3}\simeq 10^{120}.
\end{eqnarray}
We can also obtain the value of the emerged CC if we find the value of $L_\text{max}$ which is related to the CC by Eq.(\ref{Lambda}).  Since observations show that our Universe is presently entering dark energy domination, the growth of the event horizon has slowed, and it is almost as large now as it will ever become. Therefore, we can estimate the value of $L_\text{max}$ as the present value of cosmic event horizon. The present radius of the cosmic event horizon is $L_\text{CEH}\simeq  15.7\pm0.4Glyr\simeq 10^{61}L_\text{P}$ \cite{Chas}. These values together with the Eq.(\ref{Lambda}) lead us to
\begin{eqnarray}\label{1233}
\Lambda\simeq\frac{1}{L_\text{max}^2}=\frac{1}{\mathfrak{N}L^2_\text{P}}\simeq 10^{-122}M_\text{P}^2,
\end{eqnarray}
which is consistent with the observed value of the CC.

At the end of this section, let us concentrate on the coincidence problem. Cosmological observations suggest that we live in an remarkable  period in the history of the Universe when $\rho_\Lambda\simeq\rho_m$, where $\rho_\Lambda$ and $\rho_m$ are the energy density of the CC and the matter respectively.
Within the standard model of cosmology, this equality of energy densities just at the present epoch can be seen as coincidental
since it requires very special initial conditions in the very early Universe. The corresponding ``why now'' question constitutes
the cosmological ``coincidence problem''.  If  $M_\text{U}$ denotes the total mass of the pressureless matter (dust) content of the
Universe, then $M_\text{U}=m_bN_b$, where $m_b$ and $N_b$ are the mass and the total number of particles of matter content of Universe. As we know the total number of particles is approximately  equal to the entropy of the matter, $N_b\simeq S_{\text{(dust)}}$ \cite{Ama}. Also, if we use the well-known relation between the radius of the Universe (herein the present value of cosmic event horizon, $L_\text{CEH}\simeq L_\text{max}$) and mass of nucleons, $m_b$, as a result of the uncertainty principle \cite{Siv}, $m_bL_\text{max}\sim\sqrt{N_b}$, we obtain
\begin{eqnarray}\label{last10}
m_b\simeq\frac{\sqrt{N_b}}{L_\text{max}}\simeq\frac{S^\frac{1}{2}_\text{(dust)}}{\mathfrak{N}^\frac{1}{2}L_\text{P}}\simeq\frac{\mathfrak{N}^\frac{1}{3}}{\mathfrak{N}^\frac{1}{2}L_\text{P}}=\mathfrak{N}^{-\frac{1}{6}}M_\text{P}.
\end{eqnarray}
As a consequence, the total mass of Universe can be rewritten as
\begin{eqnarray}\label{last2}
\begin{array}{cc}
M_\text{U}=N_bm_b\simeq S_\text{(dust)}M_\text{P}\mathfrak{N}^{-\frac{1}{6}}=\\\mathfrak{N}^{\frac{2}{3}}\mathfrak{N}^{-\frac{1}{6}}M_\text{P}=\mathfrak{N}^\frac{1}{2}M_\text{P}.
\end{array}
\end{eqnarray}
Now we can summarize Eqs. (\ref{dis}), (\ref{Lambda}), (\ref{last10}) and (\ref{last2}) as the following scaling relations
\begin{eqnarray}\label{last3}
\begin{array}{cc}
M_\text{U}\simeq M_\text{P}\mathfrak{N}^{\frac{1}{2}},\\
m_b\simeq M_\text{P}\mathfrak{N}^{-\frac{1}{6}},\\
M_\text{UV}\simeq M_\text{P}\mathfrak{N}^{-\frac{1}{4}},\\
L_\text{max}\simeq L_\text{P}\mathfrak{N}^{\frac{1}{2}},\\
L_\text{min}\simeq L_\text{P}\mathfrak{N}^{-\frac{1}{2}},\\
\Lambda\simeq L_\text{P}^{-2}\mathfrak{N}^{-1},
\end{array}
\end{eqnarray}
which are in fact the extension of the Dirac large numbers hypothesis (LNH) explained in \cite{Mena}. Note that all of these scaling relations are established at the present time, because the observations show the entrance of Universe in the acceleration phase just at the present epoch.
Therefore,  the LNH of Dirac can actually be explained in terms of the quantum deformation of quantum Universe. Eliminating $\mathfrak{N}$ from the second and the fourth scaling relations gives us
\begin{eqnarray}\label{last4}
m_b\simeq \left(\frac{1}{GL_\text{max}}\right)^\frac{1}{3},
\end{eqnarray}
which is the empirical Weinberg formula for the mass of the nucleon \cite{Wi1}. Also, by eliminating $\mathfrak{N}$ from the second (or third one) and the last scaling relations in (\ref{last3}) we obtain
\begin{eqnarray}\label{last5}
\begin{array}{cc}
\rho_\Lambda=\frac{\Lambda}{8\pi G}\simeq M_\text{P}^4\mathfrak{N}^{-1}\simeq Gm_b^6,\\
M^4_\text{UV}\simeq M^4_\text{P}\mathfrak{N}^{-1}\simeq Gm_b^6.
\end{array}
\end{eqnarray}
These equations are identical to the scaling law proposed by Zeldovich \cite{ZZ} for the value CC. Let us now obtain the energy density of dust at the present epoch. The linear size of Universe at the present time is approximately is equal to the $L_\text{CEH}\simeq L_\text{max}$. Hence, by inserting the first, the second and the third scaling relations obtained in (\ref{last3}) into the definition of the energy density of dust, we find
\begin{eqnarray}\label{last6}
\rho_m\simeq \frac{M_\text{U}}{L_\text{max}^3}\simeq m_bM_\text{P}^3\mathfrak{N}^{-\frac{5}{6}}\simeq  Gm_b^6.
\end{eqnarray}
It is clear that the last equality is established just at the presence epoch of cosmic evolution. Therefore, Eqs. (\ref{last5}) and (\ref{last6}) show that the present values of the densities of dark energy and matter are of the same
order of magnitude, $\rho_\Lambda/\rho_m\simeq\mathcal O(1)$.

The last interesting equation which can be derived from the first scaling equation of  (\ref{last3}) is
\begin{eqnarray}\label{last7}
\mathfrak{N}\simeq 4\pi G M^2_\text{U}.
\end{eqnarray}
The right hand side of this relation is the entropy of a black hole with size of Universe. On the other hand, the left hand side, as we showed in Eq.(\ref{Hol}), represents the entropy of the holographic screen, $S_\text{dS}$. In other words,  the Universe can have no more states than that of a black hole of the same size.
}

\section{Conclusion }

In this paper, we have investigated the quantum deformation of  a spatially closed Friedmann-Lema\^itre-Robertson-Walker universe in the presence of a conformally coupled scalar field. The gravitational part of super-Hamiltonian has a self-adjoint extension if the wave function satisfies the standard Dirichlet or Neumann boundary condition.  As was shown in Ref. \cite{Bo}, there is a deep relation between the boundary conditions and the symmetries of cosmological models. In the model investigated here, the conformal invariance of the scalar field part of action functional leads us to the Heisenberg-Weyl symmetry. On the other hand, the spatial closeness of the spacetime along with the boundary conditions, mentioned previously, demands that symmetry of gravitational part is $SU(1,1)$ group. Consequently, the corresponding quantum groups of the model after deformation will be ${\mathcal U}_q(h_4)$ and ${\mathcal U}_q(su(1,1))$
quantum groups.

The quantum deformation of our cosmological model, causes the quantization of the scalar field, scale factor and the corresponding momenta. In addition, the initial Big-Bang singularity is absent in the sense that the quantized scale factor does not have the zero eigenvalue.  On the other hand, the scale factor operator is bounded from above. This means that the Universe reaches the minimum possible value of the curvature and has to stay in this state.
 Also, we show that the energy densities of dark energy (CC) and the baryonic matter are of the same order of magnitude at the present epoch of cosmic evolution. Also, the quantum deformation  causes a quantum sphere
$S_q^2$ (or equivalently a fuzzy sphere $S^2_F$) as the causal horizon with elementary cells
of Planck's area. The number of fundamental cells of the surface of a fuzzy horizon
is equals to the dimension of Hilbert space of the scalar field part of super-Hamiltonian. This allows to suggests that the CC and holographic principle are  emerged quantities as a result of the quantum deformation of quantum cosmology.
Interestingly, as it was shown in \cite{Tom}, gravitational holography is argued to render the CC stable against divergent
quantum corrections. Thus gravitational holography provides a technically natural solution to the radiatively instability of the CC \cite{Tom}.

Nevertheless, our setting here must be viewed as only one of the many attempts
trying to include quantum gravity effects into  cosmological models. It is
necessarily partial and incomplete. In order to reach more robust  conclusions
regarding e.g., the status of singularities in realistic situations, we may need to
quantize more degrees of freedom as compared to the
only two treated by us.


\vspace{1cm}

{\bf References}


\begin{thebibliography}{00}
\bibitem{Banks} T. Banks, Physics Today {\bf57}, (2004) 46 ; M.M. Sheikh-Jabbari, ``An N-tropic solution to the cosmological constant
problem'' [arXiv:hep-ph/0701084].
\bibitem{Connes1} A. Connes, {\it Noncommutative Geometry} (Academic Press, 1994).
\bibitem{Connes2}  A. Connes, M.R. Douglas, and A. Schwartz, JHEP {\bf02},  (1998) 003 [arXiv:hep-th/9711162]; N. Seiberg and E. Witten, JHEP {\bf09,} (1999) 032 [arXiv:hep-th/9908142]; T. Yoneya, Prog. Theor. Phys. {\bf103}, (2000) 1081   [arXiv:hep-th/0004074]; J. Polchinski, {\it String Theory}, Vol. 2 (Cambridge University Press, 1998).
 \bibitem{NC8} M.A. Markov, Zh. Eksp. Teor. Fiz. {\bf10}, (1940) 1313; C.N. Yang, Phys. Rev. {\bf72}, (1947) 874; H. Yukawa, Phys. Rev. {\bf91}, (1953) 415.
 \bibitem{Snyder} H.S. Snyder, Phys. Rev. {\bf71}, (1947) 38.
 \bibitem{Instanton}  N. Nekrasov and A. Schwarz, Commun. Math. Phys. {\bf198}, (1998) 689.
\bibitem{N1} J.W. Moffat, Phys. Lett. B {\bf491}, (2000) 345 ; Phys. Lett. B {\bf493}, (2000) 142.
\bibitem{N2}  A.H. Chamseddine, Phys. Lett. B {\bf504}, (2001) 33.
\bibitem{NC} H. Garcia-Compean, O. Obregon and C. Ramirez, Phys. Rev. Lett. {\bf88}, (2002) 161301  [arXiv:hep-th/0107250]; G.D. Barbosa and N. Pinto-Neto, Phys. Rev. D {\bf70}, (2004) 103512  [arXiv:hep-th/0407111]; N. Khosravi, S. Jalalzadeh and H.R. Sepangi, Gen. Rel. Grav. {\bf39}, (2007) 899 [arXiv:gr-qc/0702067]; N. Khosravi, S. Jalalzadeh and H.R. Sepangi, JHEP {\bf0601}, (2006) 134  [arXiv:hep-th/0601116];
C. Bastos, O. Bertolami, N.C. Dias and J.N. Prata, Phys. Rev. D {\bf78}, (2008) 023516  [arXiv:0712.4122].
\bibitem{Fuzzy} G. 't Hooft, Class. Quantum Grav. {\bf13}, (1996) 1023  [arXiv:gr-qc/9601014].
\bibitem{Bat}  E. Batista and S. Majid, J. Math. Phys. {\bf44}, (2003) 107  [arXiv:hep-th/0205128].
\bibitem{Bi} S. Majid and H. Ruegg, Phys. Lett. B {\bf 334}, (1994) 3.
\bibitem{Cam}  G. Amelino-Camelia, G. Gubitosi and F. Mercati, Phys. Lett. B {\bf676}, (2009) 180  [arXiv:0812.3663].
\bibitem{Q} D. Finkelstein, Phys. Rev. {\bf91}, (1969) 1261.
\bibitem{Wess} U. Carow-Watamura, M. Schlieker, M. Scholl and S. Watamura, Z. Phys. C-Particles and Fields {\bf 48}, (1990) 159.
\bibitem{Ma} S. Majid, J. Math. Phys. {\bf32}, (1991) 3246.
\bibitem{Fa}  P.P. Kulish and N. Yu. Reshetikhin, J. Sov. Math. {\bf23}, (1983) 2435; E.K. Sklyanin, Funct. Anal. Appl. {\bf16}, (1982) 263; L.D. Faddeev and L.A. Takhtajan, Springer Lect. Notes in Phys. {\bf246}, (1986) 166; L.D. Faddeev, N. Yu. Reshetikhin and L.A. Takhtajan, Alg. Anal. {\bf1}, (1987) 178.
\bibitem{Chari}  V. Chari  and A. Pressley  {\it A Guide to Quantum Groups} (Cambridge: Cambridge University Press, 1994).
\bibitem{Oh} C.H. Oh  and K.J. Singh, Phys. A: Math. Gen. {\bf25}, (1992) L149; [arXiv:hep-th/9407081].
\bibitem{Omar} O. Foda, M. Jimbo, T. Miwa, K. Miki and A. Nakayashiki, J. Math. Phys. {\bf35}, (1994) 13  [arXiv:hep-th/9305100].
\bibitem{Kaf} L.H. Kauffman,  and S.J. Lomonaco Jr., J. Knot Theory Ramifications {\bf16}, (2007) 267 .
\bibitem{Top}  C. Nayak, S.H. Simon, A. Stern, M. Freedman, and S. Das Sarma, Rev. Mod. Phys. {\bf80}, (2008) 1083  [arXiv:0707.1889].
\bibitem{Den} D. Bonatsos and C. Daskaloyannis, Prog. Part. Nucl. Phys. {\bf43}, (1999) 537  [arXiv:nucl-th/9909003].
\bibitem{Cha} Z. Chang, H.Y. Guo, and H. Yan, Phys. Lett. A {\bf156}, (1991) 192.
\bibitem{Holo} G. 't Hooft, ``Dimensional Reduction in Quantum Gravity'', [arXiv:hep-th/931026]; L. Susskind, J. Math. Phys. {\bf36}, (1995) 6377
[arXiv:hep-th/9409089].
\bibitem{q-0} L. Crane, J. Math. Phys. {\bf36}, (1995) 6180  [arXiv:gr-qc/9504038]; L. Smolin, J. Math. Phys. {\bf36}, (1995) 6417  [arXiv:gr-qc/9505028].
\bibitem{unital}L.C. Biedenharn and M.A. Lohe,  {\it Quantum Group Symmetry and q-Tensor Algebras}, (World Scientific, Singapore, 1995).
\bibitem{Chaichian}  M. Chaichian and A. Demichev, {\it Introduction to Quantum Groups}, (World Scientific, Singapore, 1996);  A. Klimyk and K. Schm\"udgen, {\it Quantum groups and their representations}, (Berlin, Springer 1997).
 \bibitem{Fin} R.J. Finkelstein,  Lett. Math. Phys. {\bf 38}, (1996) 53.
\bibitem{Cas}  L. Castellani, Phys. Lett. B {\bf 327}, (1994) 22   [arXiv:hep-th/9402033]; P. Aschieri and L. Castellani, Int. J. Mod. Phys. A {\bf 11}, (1996) 4513.
[arXiv:q-alg/9601006]; G. Bimonte, R. Musto, A. Stern and P. Vitale, Nucl. Phys. B {\bf525}, (1998) 483.
\bibitem{Com} A.Y. Alekseev, H. Grosse and V. Schomerus, Commun. Math. Phys. {\bf172}, (1995) 317 ; E. Buffenoir, K. Noui and P. Roche,  Class. Quant. Grav. {\bf19}, (2002) 4953.
\bibitem{Spin} W.J. Fairbairn and C. Meusburger, J. Math.
Phys. {\bf53}, (2012) 022501  [arXiv:1012.4784]; J.C. Baez, J.D. Christensen, T.R. Halford and D.C. Tsang,  Class. Quant. Grav. {\bf19}, (2002) 4627  [gr-qc/0202017].
\bibitem{Han} M. Han, J. Math. Phys. {\bf52} (2011) 072501 [arXiv:1012.4216]; A. Perez, Class.Quantum Grav. {\bf20} (2003) R43 [arXiv:gr-qc/0301113].
\bibitem{Po} J. Lukierski, H. Ruegg and A. Nowicky, Phys. Lett. B {\bf293}, (1992) 344; A. Ballesteros, F. J. Herranz, M. A. del Olmo and M. Santander, Phys. Lett. B {\bf351}, (1995) 137.
\bibitem{SR}  G. Amelino-Camelia, Phys. Lett. B {\bf510}, (2001) 255; J. Magueijo  and L. Smolin, Phys. Rev. Lett. {\bf88}, (2002) 190403.
\bibitem{Ya} I.Ya. Aref'eva and I.V. Volovich, Phys. Lett. B {\bf268}, (1991) 179.
\bibitem{C1}C.G. Jr. Callan, S. Coleman, and R. Jackiw, Ann. Phys. (N.Y.) 59, (1970) 42; L.H. Ford and D.J. Toms, Phys. Rev.
D {\bf25}, (1982) 1510 ; L.H. Ford, Phys. Rev. D {\bf35}, (1987) 2955.
\bibitem{C2} L. Boubekeur, E. Giusarma, O. Mena, and H. Ram\'irez, Phys. Rev. D {\bf91}, (2015) 103004  [arXiv:1502.05193].
\bibitem{Interior} S. Jalalzadeh and B. Vakili, Int. J. Theor. Phys. {\bf51}, (2012) 263 [arXiv:1108.1337].
\bibitem{Light} P. Pedram and S. Jalalzadeh, Phys. Lett. B {\bf660}, (2008) 1  [arXiv:0712.2593].
\bibitem{KK} J. Wudka, Phys. Rev. D {\bf35}, (1987) 3255.
\bibitem{MD} N. Khosravi, S. Jalalzadeh and H.R. Sepangi, Gen. Rel. Grav. {\bf39}, (2007) 899  [arXiv:gr-qc/0702067)]; N. Khosravi, S. Jalalzadeh and H.R. Sepangi,  JHEP {\bf01}, (2006) 134  [arXiv:hep-th/0601116]; S. Jalalzadeh, F. Ahmadi and H.R. Sepangi, JHEP {\bf0308}, (2003) 012  [arXiv:hep-th/0308067].
\bibitem{Ei} A. Einstein, {\it Essays in Science}, (Philosophical Library, New York, 1934).
\bibitem{Linde} A. Linde, Phys. Lett. B {\bf351}, (1995) 99.
\bibitem{TZ} E.P. Tryon, Nature {\bf246}, (1973) 396; Ya.B. Zel'dovich, Sov. Astron. Lett. {\bf7}, (1981) 322;  J.J. Ferrando, R. Lapiedra, and J.A. Morales, Phys. Rev. D {\bf75}, (2007) 124003.
\bibitem{Lan} L. Landau and E. M. Lifshitz, {\it The Classical Theory of Fields} (Elsevier, Amsterdam, Fourth
ed., 1975; Reprinted in 2007).
\bibitem{Haw} J.B. Hartle and S.W. Hawking, Phys. Rev. D {\bf28}, (1983) 2960; J.J. Halliwell and S.W. Hawking, Phys. Rev. D {\bf31}, (1985)
1777.
\bibitem{Ver} A. Vilenkin, Phys. Lett. B {\bf117}, (1982) 25; A. Vilenkin, Phys. Rev. D {\bf27}, (1983) 2848 ; A.D. Linde, Lett. Nuovo Cimento
Soc. Ital. Fis. {\bf39}, (1984) 401.
\bibitem{De} B.S. De Witt, Phys. Rev. {\bf160}, (1967) 1113.
\bibitem{Mosh} V.G. Lapchinskii and V. A. Rubakov, Theor. Math. Phys. {\bf33}, (1977) 1076 ; N. A. Lemos, J. Math. Phys. (N.Y.) {\bf37}, (1996) 1449.
\bibitem{St} V.A. Strauss and M.A. Winklmeier, Vestnik YuUrGU. Ser. Mat. Model. Progr., {\bf 9}, (2016) 73.
\bibitem{Neumann}  J. von Neumann, {\it Mathematical Foundations of Quantum Mechanics} (Princeton University Press, Princeton, 1955).
\bibitem{Franco} F. Strocchi, {\it An Introduction to the Mathematical Structure of Quantum Mechanics}, 2nd Ed., (World Scientific, Singapore, 2008)
\bibitem{Derin} V.G. Drinfel'd, J. Math. Sci. {\bf41}, (1988) 898.
\bibitem{Alg} E. Abe, {\it Hopf Algebras}, Cambridge tracs in nathematics 74, Cambridge; Univ. Presss (1980).
\bibitem{Ku} P. Kulish and E. Damaskinsky, J. Phys. A {\bf23}, (1990) L415.
\bibitem{Bon}  D. Bonatsos et al., Phys. Lett. B {\bf331}, (1994) 150  [arXiv: hep-th/9402014].
\bibitem{Her} G. Szeg\H o, {\it Orthogonal Polynomials}, 4th ed., Amer. Math. Soc. Colloquium Publications,
vol. 23, Amer. Math. Soc., Providence, R.I., (1975).
\bibitem{Dama} E.V. Damaskinskii and P.P. Kulish,  J. Math. Sci. {\bf62}, (1992) 2963.
\bibitem{La} L. Gatteschi, J. Comput. Appl. Math. {\bf144}, (2002) 7.
\bibitem{Loop} K. Noui, A. Perez, and D. Pranzetti, JHEP {\bf10}, (2011) 36  [arXiv:1105.0439]; D. Pranzetti, Phys. Rev. D {\bf89}, (2014) 084058  [arXiv:1402.2384]; W.J. Fairbairn and C. Meusburger, J. Math.
Phys. {\bf53}, (2012) 022501  [arXiv:1012.4784]; H.M. Haggard, M. Han, W. Kami\'{n}ski, A. Riello, Phys Lett. B {\bf752},  (2016) 258 [arXiv:1509.00458].
\bibitem{Du} M. Dupuis and F. Girelli, Phys. Rev. D {\bf90}, (2014) 104037  [arXiv:1311.6841].
\bibitem{Wi} S. Weinberg, Rev. Mod. Phys., 61 (1989) 1; T. Padmanabhan,
Cosmological constant-The weight of the vacuum [hep-th/0212290], and references therein.
\bibitem{Bur} C.P. Burgess, “The Cosmological Constant Problem: Why it's hard to get Dark Energy from Microphysics,”
[arXiv:1309.4133].
\bibitem{Li} M. Li, X. D. Li, S. Wang and Y. Wang, Dark energy, Commun. Theor. Phys. {\bf56} (2011) 525
[arXiv:1103.5870].
\bibitem{J} E. Bianchi and C. Rovelli, Phys. Rev. D, {\bf84}, (2011) 027502 [arXiv:1105.1898].
\bibitem{Sphere}  P. Podles, Lett. Math. Phys. {\bf14}, (1987) 193 .
\bibitem{Mad} H. Grosse, J. Madore and H. Steinacker, J. Geom. Phys. {\bf38}, (2001) 308  [arXiv:hep-th/0005273].
\bibitem{F} J. Madore,  Class. Quantum Grav. {\bf9}, (1992) 69.
\bibitem{Euc} Y. Abe and V.P. Nair, Phys. Rev. D {\bf68}, (2003) 025002.
\bibitem{Sheykh} M.M. Sheikh-Jabbari, Phys. Lett. B, {\bf642}, (2006) 119  [arXiv:hep-th/0605110].
\bibitem{Martin}  J. Martin, Comptes Rendus Physique {\bf13} (2012) 566 [arXiv:1205.3365].
\bibitem{Chas} C.A. Egan and C.H. Lineweaver, , Astroph. J.{\bf710}, (2010) 1825 [arXiv:0909.3983].
\bibitem{Ama} M. Amarzguioui and O. Gr{\o}n, Phys. Rev. D {\bf71}, (2005) 083011 [arXiv:gr-qc/0408065].
\bibitem{Siv} C. Sivaram, Astrophys. Space Sci. {\bf124}, (1986) 195.
\bibitem{Mena} G.A. Mena Marugan and S. Carneiro, Phys. Rev. D {\bf65}, (2002) 087303 [arXiv:gr-qc/0111034].
\bibitem{Wi1} S. Weinberg, Gravitation and Cosmology (Wiley, New York, 1972).
\bibitem{ZZ} Y.B. Zeldovich, Sov. Phys. Usp, {\bf11}, 381 (1968); Y.B. Zeldovich,
Sov. Phys. Usp, {\bf24}, 216 (1981).
\bibitem{Bo} S. Jalalzadeh and P.V. Moniz, Phys. Rev. D {\bf89}, (2014)  083504  [arXiv:1403.2424];
S. Jalalzadeh, T. Rostami and P.V. Moniz, Eur. Phys. J. C {\bf75}, (2015) 38  [arXiv:1412.6439]; T. Rostami, S. Jalalzadeh and P.V. Moniz, Phys. Rev. D. {\bf92}, (2015) 023526  [arXiv:1507.04212]; S. Jalalzadeh, T. Rostami and P.V. Moniz, Int. J. Mod. Phys. D {\bf25}, (2016) 1630009.
\bibitem{Tom} S. Thomas, Phys. Rev. Lett. {\bf89}, (2002) 081301.
\end{thebibliography}
\end{document}